\documentclass[superscriptaddress,aps,pra,a4paper,nofootinbib,preprint]{revtex4}
\usepackage{amsfonts,amsmath,amssymb}
\usepackage{bm,tensor}
\usepackage{graphicx}
\usepackage{braket}
\usepackage{ifpdf}
\usepackage[utf8]{inputenc}
\usepackage{color}
\usepackage{ulem}
\usepackage{comment}
\usepackage[colorlinks=true
,urlcolor=blue
,anchorcolor=blue
,citecolor=blue
,filecolor=blue
,linkcolor=blue
,menucolor=blue
,linktocpage=true
,pdfproducer=medialab
,pdfa=true
]{hyperref}
\newcommand{\df}{\text{d}}
\newcommand{\p}{\partial}
\newcommand{\Mpl}{{M_\text{pl}}}

\begin{document}

\begin{flushright}
RESCEU-23/21
\end{flushright}

%%%%%%%%%%%%     title
\title{
Gauge Field Production and Schwinger Reheating in Runaway Axion Inflation
}

\author{Soichiro Hashiba}
\affiliation{Department of Physics, Graduate School of Science,
The University of Tokyo,\\ Hongo 7-3-1
Bunkyo-ku, Tokyo 113-0033, Japan}
\affiliation{Research Center for the Early Universe (RESCEU), Graduate School of Science,\\ The University of Tokyo, Hongo 7-3-1, Bunkyo-ku, Tokyo 113-0033, Japan}

\author{Kohei Kamada}
\affiliation{Research Center for the Early Universe (RESCEU), Graduate School of Science,\\ The University of Tokyo, Hongo 7-3-1, Bunkyo-ku, Tokyo 113-0033, Japan}

\author{Hiromasa Nakatsuka}
\affiliation{Department of Physics, Graduate School of Science,
The University of Tokyo,\\ Hongo 7-3-1
Bunkyo-ku, Tokyo 113-0033, Japan}
\affiliation{Institute for Cosmic Ray Research (ICRR), The University of Tokyo,\\ Kashiwa-no-ha 5-1-5, Kashiwa-shi, Chiba 277-8582, Japan}

\begin{abstract}

In a class of (pseudoscalar) inflation, inflationary phase is followed by a kination phase, 
where the Universe is dominated by the kinetic energy of the inflaton that runs away in a vanishing scalar potential. 
In this class of postinflationary evolution of the Universe, reheating of the Universe cannot be achieved by the inflaton 
particle decay, which requires its coherent oscillation in a quadratic potential. 
In this study, we explore the U(1) gauge field production through the Chern-Simons coupling between the pseudoscalar
inflaton and the gauge field during the kination era
and examine the subsequent pair-particle production induced by the amplified gauge field known as the Schwinger effect, 
which can lead to reheating of the Universe. 
We find that with a rough estimate of the Schwinger effect for the Standard Model hyper U(1) gauge field 
and subsequent thermalization of the pair-produced particles, 
a successful reheating of the Universe can be achieved
by their eventual domination over the kinetic energy of the inflaton,
with some reasonable parameter sets. 
This can be understood as a concrete realization of the ``Schwinger reheating''. 
Constraints from the later-time cosmology are also discussed.

\end{abstract}

\maketitle
\date{\today}
\tableofcontents

%%%%%%%%%%%%     main document 

\section{Introduction}

Inflation driven by a pseudoscalar or an axion-like particle (ALP), also often dubbed as natural inflation~\cite{Freese:1990rb,Adams:1992bn},
is one of the most well-motivated models of inflation.
While one of the difficulties in constructing inflation models is how to realize a flat potential suitable for inflation 
against the quantum corrections, 
in axion inflation the flatness of the potential is guaranteed by the shift symmetry,
which is the nature of the (pseudo) Nambu-Goldstone bosons.
On the one hand, it is difficult to drive inflation by the original QCD axion~\cite{Weinberg:1977ma,Wilczek:1977pj,
Kim:1979if,Shifman:1979if,Zhitnitsky:1980tq,Dine:1981rt}
due to the requirements on its potential as well as other interactions as the solution for the strong CP 
problem~\cite{Peccei:1977hh,Peccei:1977ur}.  
On the other hand, ALPs that arise from, {\it e.g.}, superstring theory compactifications~\cite{Svrcek:2006yi}, is allowed to have 
a non-trivial potential to drive the inflation that fits the observational data, 
as is seen in the axion monodromy~\cite{Silverstein:2008sg,McAllister:2008hb}.
Moreover, ALPs can be applied for the model building of the inflation 
with an extremely flat potential such as $k$-inflation~\cite{Armendariz-Picon:1999hyi} or the quintessential inflation~\cite{Peebles:1998qn}, 
since the shift symmetry can explain the flatness of the potential required in these models.

While inflation models with such an extremely flat potential and a resultant cosmic history 
have interesting phenomenologies, 
the connection to the hot Big Bang Universe is not clear. 
In such models, after inflation the inflaton runs away or continue to move in a vanishing potential. 
The energy density of the Universe is then dominated by the kinetic energy of the inflaton, 
which is often dubbed as the kination era~\cite{Spokoiny:1993kt,Joyce:1996cp}. 
Since the inflaton does not oscillate coherently around the potential minimum, 
reheating of the Universe cannot be achieved by the inflaton particle decay.
Instead, reheating is achieved by a 
small amount of radiation produced at a time after inflation that eventually dominates the energy density of the Universe. 
Note that the inflaton
kinetic energy {during the kination era} is damped by the cosmic expansion as $\propto a^{-6}$, 
{with $a$ being the scale factor, which is}  faster than {that of} radiation.
Consequently, the production of the small amount of radiation is the key ingredient 
for the graceful exit in these models.

One of the most representative mechanisms to produce a small amount of radiation after inflation 
before or during kination is the gravitational particle production~\cite{Parker:1969au,Zeldovich:1971mw,Ford:1986sy}, 
where all the non-conformally coupled fields are produced due to the change 
of the vacuum caused by the change of the background spacetime at the end of inflation.
This ``gravitational reheating'' is, however, {not so efficient} due to the weakness of gravity 
and often suffered from too much production of high-frequency gravitational waves, 
which are also produced gravitationally.
The latter is constrained by the number of relativistic degrees of freedom at the Big Bang Nucleosynthesis (BBN)~\cite{Fields:2019pfx} 
and at the recombination with the observation of the cosmic microwave background (CMB)~\cite{Aghanim:2018eyx}. 
Although there are intensive studies to address this issue 
and to find successful scenarios in the gravitational reheating~\cite{Pallis:2005bb,Chun:2009yu,Kunimitsu:2012xx,DeHaro:2017abf,AresteSalo:2017lkv,Hashiba:2018iff,Hashiba:2018tbu,Nakama:2018gll,Herring:2020cah,Babichev:2020yeo,Hashiba:2019mzm,Lankinen:2019ifa,Lankinen:2020mrc}, 
it is important to explore other possibilities of reheating in the inflationary models with kination.

The difficulty in producing even small amount of radiation lies in the fact that
the shift symmetry forbids or at least suppresses ordinary couplings of the inflaton to other fields.
However, once we identify that the inflaton is an ALP, a nontrivial coupling between the inflaton and other fields 
that respects the shift symmetry, that is, the Chern-Simons coupling, naturally arises. 
The Chern-Simons coupling is induced  {\it e.g.}, if the underlying Peccei-Quinn-like global symmetry 
is anomalous under some local gauge symmetries 
or by the Green-Schwarz mechanism of anomaly cancellation~\cite{Green:1984sg} in heterotic string theory. 
If the gauge symmetry is a U(1) symmetry, a coherent axion dynamics is known to 
produce the U(1) gauge fields copiously through 
their tachyonic instability~\cite{Turner:1987bw,Garretson:1992vt,Anber:2006xt}.  
These phenomena during inflation
have been studied to constrain the strength of the coupling through the cosmological observations~\cite{Barnaby:2010vf,Sorbo:2011rz,Barnaby:2011vw,Barnaby:2011qe,Meerburg:2012id,Pajer:2013fsa,Linde:2012bt}. 
Moreover, if the U(1) gauge field is the one in the Standard Model of particle physics (SM),
it can also explain the baryon asymmetry of the Universe~\cite{Anber:2015yca,Jimenez:2017cdr,Domcke:2018eki,Domcke:2019mnd} and the origin 
of the intergalactic magnetic fields~\cite{Neronov:2010gir,Tavecchio:2010mk,Dolag:2010ni,Finke:2015ona,Fermi-LAT:2018jdy} 
(See also the studies on the gauge field amplification during reheating~\cite{Fujita:2015iga,Adshead:2016iae,Cuissa:2018oiw,Okano:2020uyr,Brandenburg:2021bfx}). 
It is natural that we expect that it would also lead to a successful reheating after kination, 
if the tachyonic instability of the U(1) gauge fields is sufficiently effective during kination.

In this article, we study the U(1) gauge field amplification triggered by the Chern-Simons coupling during kination.
We take an inflation model with a ``runaway''-type or a step-function-like potential as an example.
In this kind of model, the gauge field amplification during kination is more efficient than during inflation
since the ALP velocity takes its maximal value just at the onset of kination when 
almost all the potential energy is converted into the kinetic energy of the inflaton. 
We can easily see that due to the smallness of the kinetic energy during inflation, 
the gauge field amplification during inflation is negligibly small. 
Compared to the gauge field amplification from the ALP oscillation~\cite{Fujita:2015iga,Adshead:2016iae,Cuissa:2018oiw}, 
that from this ``runaway'' ALP is distinctive in the point that purely only one helicity modes are amplified
because the sign of the ALP velocity is unchanged, and the resultant gauge fields are maximally helical. 
We find that gauge field amplification during kination 
occurs when the mode that exited the horizon during inflation reenters it. 
It is contrasting to the one during inflation which occurs when the mode exits the horizon~\cite{Turner:1987bw,Garretson:1992vt,Anber:2006xt,Jimenez:2017cdr}. 
The energy density of the amplified gauge fields is found to be typically much larger than that
of gravitationally produced particles, which is an advantage for successful reheating.

To see if the gauge field amplification by the Chern-Simons coupling leads to the successful reheating, 
we need to examine the interactions of the SM particles. 
Indeed, even during the gauge field amplification, strong electric fields would 
induce a pair production of charged particles~\cite{Garriga:1994bm,Frob:2014zka,Cai:2014qba,Kobayashi:2014zza,Stahl:2015gaa,Bavarsad:2016cxh,Hayashinaka:2016dnt,Hayashinaka:2016qqn,Hayashinaka:2018amz}, known as the Schwinger effect~\cite{Schwinger:1951nm,Gelis:2015kya}, 
which also backreacts to the gauge-field dynamics. 
Unfortunately, the Schwinger effect  caused by a dynamical gauge field is extremely difficult to give a precise estimate
with the best of our present knowledge and technique, 
although there are several trials to challenge this problem~\cite{Domcke:2018eki,Sobol:2019xls,Gorbar:2021rlt}. 
In this article, we adopt the treatment developed in Ref.~\cite{Domcke:2018eki} to give a rough estimate. 
Although it would not give a precise estimate and the results in the present paper are not quantitatively correct, 
we believe that the estimate obtained in this way gives us the characteristic 
feature of the entire process of the gauge field amplification during kination. 
We find that the energy density of the pair-produced particles is typically as much as or even larger than 
that of the gauge fields
and they eventually thermalized before dominating the energy density of the Universe. 
Therefore, the successful reheating through the Schwinger effect, or the ``Schwinger reheating''~\cite{Tangarife:2017rgl} is realized. 
The electric fields are likely to be screened just after the saturation of the gauge field amplification, 
while the magnetic fields would decay slowly. 
The latter can lead to a Universe inconsistent with the present one due to the too much additional number of relativistic
degrees of freedom, parameterized by the number of effective neutrino species, $N_\mathrm{eff}$, 
constrained by the BBN~\cite{Fields:2019pfx}. 
It turns out that taking into account the magnetohydrodynamics (MHD)  
cascade decay, the energy density of the magnetic fields 
are diffused enough by the time of the BBN and $N_\mathrm{eff}$ is sufficiently suppressed.

The rest of this paper is constructed in the following way.
In Sec.~\ref{sec_gaugefieldproduction}, we examine the gauge field production during the kination era 
in the ALP-photon system without other charged particles
and derive the approximate formula for its energy density.
In Sec.~\ref{sec_schwinger}, we include charged fermions to give a rough estimate for the Schwinger effect
including the backreaction on the gauge field production.
We also investigate the thermalization of the produced particles and screening of the electric field 
after the saturation of the gauge field amplification.
In Sec.~\ref{sec_timeevolution}, we examine the late time evolution of magnetic fields {and discuss cosmological constraints}.
Sec.~\ref{sec_conclusion} is devoted for conclusion and discussion.

%%%%%%%%%%%%%%%%%%%%%%%%%%%%%%%%%%%%%%%%%%%%
\section{Gauge field amplification in runaway-type inflation}
\label{sec_gaugefieldproduction}
%%%%%%%%%%%%%%%%%%%%%%%%%%%%%%%%%%%%%%%%%%%%

Let us start with investigating the gauge field production in inflationary scenarios with the kination era without taking into account the backreaction
from the Schwinger effect.
We focus on the gauge field dynamics by the motion of the ALP field, $\phi$, which acts as the inflaton with the potential $V(\phi)$.
We consider the following action in the Friedmann-Robertson-Walker (FRW) background,
\begin{align}
    \mathcal S &= \int d^4 x  \sqrt{-g} \left[
    \mathcal L_{\rm \phi}
    +\mathcal L_{\rm EM}
    +\mathcal L_{\rm CS}
    + \mathcal L_\psi \right], 
    \label{eq_lagrangian}
\\
    \mathcal L_{\rm \phi}
	 &= -\frac{1}{2}\p_\mu \phi \p^\mu \phi 
	- V(\phi),
    \label{eq_lagrangian_phi}
\\
    \mathcal L_{\rm EM}
	 &= 
	-\frac{1}{4} F_{\mu\nu} F^{\mu\nu} ,
	\label{eq_lagrangian_EM}
\\
    \mathcal L_{\rm CS}
    &=
	- \frac{ \phi}{4\Lambda } F_{\mu\nu} \tilde F^{\mu\nu}, 
	\label{eq_lagrangian_CS}
\\
    \mathcal L_\psi &= i\bar\psi \gamma^\mu D_\mu \psi = i\bar\psi \left[ a^{-1} \gamma^\mu \left( \partial_\mu +ig'Q A_\mu \right) + \frac32 H\gamma^0 \right] \psi, 
    \label{eq_lagrangian_fermi}
\end{align}
where $F_{\mu\nu}\equiv \partial_\mu A_\nu- \partial_\nu A_\mu$ is the (hypercharge) U(1) gauge field strength, $\tilde F^{\mu\nu} \equiv \epsilon^{\mu\nu\rho\sigma} F_{\rho\sigma} /(2\sqrt{-g})$ is its dual, and $\gamma^\mu$ is the gamma matrix satisfying $\{ \gamma^\mu, \gamma^\nu \} = -2 \eta^{\mu\nu}$.
$\mathcal L_{\rm CS}$ is the Chern--Simons term, which is the key ingredient to amplify the gauge field. $\Lambda$ denotes an effective suppression scale, whose amplitude is related to the scale at which 
the anomalous coupling is generated. In this article, we take it as a free parameter.
$\mathcal L_\psi$ with the charged fermion, $\psi$, is introduced 
for Sec.~\ref{sec_schwinger} where we will 
examine the 
particle production from the strong gauge field or the Schwinger effect.
$g'$ is the (hyper) U(1) gauge coupling, and as the reference value we take it as 0.3. 
$Q$ is the (hyper)charge of the $\psi$ field.
$H \equiv {\dot a}/a$ is the Hubble parameter with the dot being the derivative with respect to the physical time $t$ and $a(t)$ denoting the scale factor in the FRW metric $ds^2 =-dt^2 + a^2(t) d{\bm x}^2 = - a^2(\eta)(d\eta^2-d{\bm x}^2)$
($\eta$ is the conformal time).
Since we have in mind the case where the mechanism works at a higher temperature than the electroweak scale, we identify the U(1) gauge field is the hyper U(1) gauge field in the standard model later. 
We however do not distinguish the hyper U(1) gauge field
and the electromagnetic U(1) gauge field otherwise stated, since the discussion does not change.

We investigate a ``runaway-type'' inflation where the inflaton energy is converted  mostly into the kinetic energy of the inflaton at the end of inflation so that the inflaton ``runs away'' in a flat direction of a vanishing potential. 
It is realized, for example, by a step-function like potential with a mild tilt where
the potential has a flat region with a non-vanishing potential energy for slow-roll inflation
and another flat region with a vanishing potential energy for the runaway. 
For concreteness, we here have
the following toy-model potential in mind,
\begin{align}
V(\phi) =
    3 \Mpl^2 H_I^2  (1-\theta(\phi)), \label{eq_toymodel}
\end{align}
where $\theta$ is the unit step function, which describes the steep cliff at $\phi=0$. 
$H_I$ is the Hubble parameter during inflation, and $\Mpl$ is the reduced Planck mass.  
Inflation takes place at $\phi<0$ and the runaway reheating stage takes place at $\phi>0$.
We do not write the ``inclination'' term explicitly
but assume that there is a very gentle slope enough for the inflaton to keep slow-rolling with $\dot\phi>0$. 
Although we take a concrete model here, within our simplifications the essence of the  phenomena we study 
do not depend on the detail of the potential.  
During inflation, the conformal time is given in terms of the scale factor as
\begin{align}
	\eta \simeq -\frac{1}{aH_I}.
\end{align}
After inflation, the Universe enters the so-called kination era~\cite{Spokoiny:1993kt,Joyce:1996cp},  
where the energy density of the Universe is dominated by the kinetic energy of inflaton and decreases at a rate proportional to $a^{-6}$.
The conformal time after inflation is then written as 
\begin{align}
	\eta &\simeq \frac{1}{2a H(\eta)} -
	 \frac{3}{2a_{\rm end}H_I} 
	 , \label{eq_etakin}
\end{align}
so that at the end of inflation it is given by $\eta_\mathrm{end} = -1/(a_\mathrm{end}H_I)$ with $a_{\rm end}$ being the scale factor at the end of inflation. 
Since the kinetic energy of the inflaton  decays faster than that of radiation or matter, 
a small amount of them produced at, for example, the end of inflation 
eventually dominates the Universe, which is the reheating mechanism in this scenario. 
The production of such a small amount of radiation or matter after inflation in the runaway inflation scenario
is  the main topic of the present paper.

%%%%%%%%%%%%%%%%%%%%%%%%%%%%%%%%%%%%%%%%%%%%
\subsection{Gauge field amplification}
\label{sec_gaugefieldproduction_perturb}
%%%%%%%%%%%%%%%%%%%%%%%%%%%%%%%%%%%%%%%%%%%%

Let us study the dynamics of the ALP and gauge field during the inflation and the kination era.
Adopting the radiation gauge, $A_0 = \p_i A^i=0$,  the physical electric and magnetic fields are given by 
\begin{align}
	\boldsymbol E_\mathrm{p} = -\frac{1}{a^2} \boldsymbol A' 
	\quad ,\quad
	\boldsymbol B_\mathrm{p} = \frac{1}{a^2} \boldsymbol \nabla \times \boldsymbol A,  \label{ebfields}
\end{align}
where the prime denotes the derivative with respect to the conformal time $\eta$ and the subscript p represents
that the variables are evaluated in the physical frame.
The quantization of the gauge fields are performed in the momentum space with the mode functions $A_\pm(k,\eta)$ 
as
\begin{align}
A_i(t,\boldsymbol x) 
	=
	\sum_{\lambda =\pm}
	\int\frac{\df^3 k}{(2\pi)^3}
	e^{i\boldsymbol k\cdot \boldsymbol x}
	e^{(\lambda)}_i(\hat {\boldsymbol k})
	\left[
		a_{\boldsymbol k}^{(\lambda)}  A_\lambda (k,t)
		+
		a_{-\boldsymbol k}^{(\lambda)\dag }  A_\lambda^* (k,t)
	\right]. 
\end{align}
Here $\lambda=\pm$ represents the circular polarization states and 
$e^{(\lambda)}_i(\hat {\boldsymbol k})$ denotes the circular polarization vector that satisfies
\begin{equation}
e^{(\lambda)}_i(\hat {\boldsymbol k}) e^{(\lambda') *}_i(\hat {\boldsymbol k}) = \delta^{\lambda \lambda'}, \quad k^i e^{(\lambda)}_i(\hat {\boldsymbol k})=0,\quad i \epsilon^{ijk} k_j e^{(\lambda)}_k(\hat {\boldsymbol k}) = \lambda k e^{(\lambda)}_i(\hat {\boldsymbol k}),  \label{polarizationtensor}
\end{equation}
with $k \equiv |{\bm k}|$. $a_{\boldsymbol k}^{(\lambda)}  $ and $a_{\boldsymbol k}^{(\lambda)\dag } $ are the 
annihilation and creation operators for the state $|{\bm k}, \lambda \rangle$, which satisfy the usual 
commutation relations, $[a_{\boldsymbol k}^{(\lambda)}  , a_{{\boldsymbol k}'}^{(\lambda')\dag } ]=\delta^{\lambda \lambda'} \delta^{(3)} ({\bm k} - {\bm k}')$. 
If only one circular polarization mode exists, from Eqs.~\eqref{ebfields} and \eqref{polarizationtensor}, 
one can see that the electric and magnetic fields are effectively parallel.

The equation of motion for the homogeneous mode of the ALP, $\phi_0$, and the mode equation for the gauge fields derived from Eq.~\eqref{eq_lagrangian} are given as
\begin{align}
			\ddot \phi_0 (t)
			+3H\dot\phi_0(t)
			+\frac{\partial V}{\partial \phi}
			&=\frac{1}{\Lambda} \braket{\boldsymbol E_\mathrm{p}\cdot \boldsymbol B_\mathrm{p}}
			\label{eq_eom_alp}
			,
	\\
			\left(
				 \p^2_\eta  
				+k^2
				\mp 2 k \xi aH
			\right)
			 A_\pm (k,\eta) 
			 &= 0, 
			 \label{eq_eom_EM}
\end{align}
where $\xi$ is the {\it instability parameter} defined as
\begin{align}
	\xi = \frac{ \dot\phi_0}{2 \Lambda H}.
	\label{eq_def_xi}
\end{align}
The angle bracket represents the quantum mechanical expectation value, 
which is identified as the classical ensemble average if it is exponentially amplified.
For concreteness, we here take $\xi>0$, 
but the gauge field amplification itself in the case with $\xi<0$ can be investigated in the same way. 
The difference is that just the opposite helicity modes are amplified. 

It would be desirable if we can simultaneously solve the equations of motion (Eqs.~\eqref{eq_eom_alp} and \eqref{eq_eom_EM}) consistently,
but it is practically difficult and model dependent. Instead we simplify the situation as follows. 
Since we consider the runaway type dynamics of the ALP as the inflaton, 
namely, the slow-roll inflation followed by the kination era, 
the ALP dynamics, $\dot\phi_0$, during inflation is suppressed by the slow roll parameter $\epsilon$ so that
we can express as
\begin{align}
	\dot\phi_0^2  = 2 \epsilon \Mpl^2 H^2 
	\quad
	\text{(during inflation)}.
\end{align}
During the kination era, on the other hand, the kinetic energy of the ALP, ${\dot \phi}_0^2/2$,  dominates the 
energy density of the Universe so that we have 
\begin{align}
	\dot\phi_0^2 = 2\Mpl^2 H^2
	\quad
	\text{(during kination)}.
\end{align}
The Hubble parameter approximately evolves as
\begin{equation}
H = \left\{\begin{array}{ll} H_I, &\quad \text{for} \quad \eta<\eta_\mathrm{end}, \\ H_I \left(\dfrac{a(\eta)}{a_\mathrm{end}}\right)^{-3}, & \quad \text{for} \quad \eta>\eta_\mathrm{end} ,
\end{array}\right.
\end{equation}
with assuming the energy density of the amplified gauge fields (and possibly generated other particles) 
is negligible.
The instability parameters (Eq.~\eqref{eq_def_xi}) in each era are then given by
\begin{align}
	\xi_I  &= \epsilon \frac{1}{\sqrt{2} } \frac{\Mpl}{\Lambda}
	\quad ,\quad
		\xi_K  =   \frac{1}{\sqrt{2} } \frac{\Mpl}{\Lambda}
		,
	\label{eq_xis}
 \end{align}
where the subscripts $I$ and $K$ mean that the quantities are evaluated during inflation and kination, respectively. 
One can see that we always have $\xi_I \ll \xi_K$. 
Note that the $\xi_I$ at the cosmic microwave background (CMB)  scale is severely constrained by the observations. For example, 
a large non-Gaussianity on the temperature perturbation of the cosmic microwave background (CMB) induced by the amplified gauge fields puts a constraint as
$|\xi_I|<2.37$~ \cite{Barnaby:2011vw,Meerburg:2012id,Pajer:2013fsa}.
In our setup, we can choose a
relatively large $\xi_K$ while keeping $\xi_I$ within the acceptable range from the observational constraints, which makes the gauge field production more efficient.

Next we solve the mode equation for the gauge fields (Eq.~\eqref{eq_eom_EM}) 
on top of the background solutions with Eq.~\eqref{eq_xis}.
During inflation, Eq.~\eqref{eq_eom_EM} is rewritten as 
\begin{align}
	&\left(
	\p^2_\eta 
	+k^2
	\pm 2 \frac{k \xi_I }{\eta}
	\right)
	 A_{I,\pm} (k,\eta) = 0, 
\end{align}
so that its general solution is given by
\begin{align}
	 A_{I,\pm} (k,\eta) 
	&=
	C_{I1}^{\pm} W_{\mp i\xi_I,1/2}(+2i k\eta)
	+C_{I2}^{\pm} W_{\pm i\xi_I,1/2}(-2i k\eta), 
\end{align}
where $W_{\kappa,\mu}(x)$ is the Whittaker function and we have taken $\xi_I$ to be a constant.
Strictly speaking $\xi_I$ is time-dependent, but it varies only slowly with time due to the slow-roll motion of 
the ALP field. 
For the practical purpose to solve Eq.~\eqref{eq_eom_EM}, it is enough to take it as a constant. 
Requiring the mode functions are taken in the Bunch-Davies-like vacuum with the asymptotic form 
$\lim_{-k \eta \rightarrow \infty}  A_{I,\pm} (k,\eta)  \sim \exp(-ik\eta)/\sqrt{2k}$,
we obtain the positive frequency mode function 
with $C_{I1}^\pm = e^{\pm \pi \xi_I/2}/\sqrt{2k}$ and $C_{I2}^\pm =0$
as~\cite{Anber:2006xt,Jimenez:2017cdr} 

\begin{align}
		 A_{I,\pm} (k,\eta)  
		&=
		\frac{1}{\sqrt{2k}} 
		e^{\pm \pi \xi_I/2}
		W_{\mp i \xi_I,1/2}(+2i k\eta), \label{solinf}
\end{align}
up to the phase factor.  
With $\xi_I>0$, 
the positive helicity modes are amplified exponentially around the horizon crossing 
whereas the negative helicity modes remain oscillatory. 
This is because the positive helicity modes are tachyonic for $k< 2 \xi_I/\eta$ 
while the effective mass squared of the negative helicity modes is always positive.

The equation of motion during the kination era is rewritten as 
\begin{align}
	&\left(
	\p^2_{\eta_{\small K}}
	+k^2
	\mp  \frac{k \xi_K }{\eta_K}
	\right)
	 A_{K,\pm} (k,\eta_K) = 0 & \text{with}& \quad \eta_K \equiv \eta+\frac{3}{2 a_\mathrm{end} H_I}, 
\end{align}
whose general solution is given by
\begin{align}
	 A_{K,\pm} (k,\eta_K) 
	&=
	C_{K1}^{\pm} W_{\pm i\xi_K/2,1/2}(+2i k\eta_K)
	+C_{K2}^{\pm} W_{\mp i\xi_K/2,1/2}(-2i k\eta_K).
	\label{eq_generalSolKination}
\end{align}
The coefficients $C_{K1}^{\pm}$ and $C_{K2}^{\pm}$ are determined by the matching conditions to the solution during inflation
Eq.~\eqref{solinf}
at the end of inflation as
\begin{align}
\begin{cases}
	 A_{I,\pm} (k,\eta_\mathrm{end}) =  A_{K,\pm} \left(k,\eta_\mathrm{end} +{\displaystyle \frac{3}{2 a_\mathrm{end} H_I}} \right) \\
	\left.\p_\eta A_{I,\pm} (k,\eta)\right|_{\eta = \eta_\mathrm{end}} = \left.\p_\eta A_{K,\pm} 
	\left(k,\eta +{\displaystyle \frac{3}{2 a_\mathrm{end} H_I}} \right) \right|_{\eta = \eta_\mathrm{end}}
\end{cases}. 
	\label{eq_determine_C12}
\end{align}
Once more, for $\xi_K>0$, regardless of the coefficients $C_{K1}^{\pm}$ and $C_{K2}^{\pm}$, 
only positive helicity modes are exponentially amplified while the negative ones remain oscillatory. 
Thus hereafter we focus on the positive helicity modes.
Moreover, as we have seen $\xi_I \ll \xi_K$ in our setup, in the following
we approximate as $\xi_I=0$ 
and investigate the case with $\xi_K > 1$. 

From the matching conditions~\eqref{eq_determine_C12} we find
\begin{align}
C_{K1}^+ &=\frac{-1}{8\sqrt{2 k}} (e^{\pi \xi_K}-1) \Gamma \left[-\frac{i \xi_K}{2} \right]\left(2 \log \left[ u\right]+2 \psi\left(\frac{i \xi_K}{2} \right) +4 \gamma_E - i \pi \right) \left(1+{\cal O}\left(u\right)\right), \label{c1match}\\
C_{K2}^+ &= \frac{-e^{\pi \xi_K/2}}{2\sqrt{2k} \Gamma \left[-\frac{i \xi_K}{2} \right]} \left( 2 \log \left[ u\right]+ 2 \psi\left(-\frac{i \xi_K}{2} \right) +4 \gamma_E+i \pi +\frac{4 i }{\xi_K}\right)\left(1+{\cal O}\left(u\right)\right),\label{c2match}
\end{align}
where $u\equiv k/(a_\mathrm{end} H_I)$, 
$\psi(z)$ is the polygamma function and $\gamma_E$ is the Euler's gamma. 
For large $\xi_K$, we find 
\begin{equation}
|C_{K1}| \simeq |C_{K2}| \simeq \frac{1}{2} \sqrt{\frac{\xi_K}{2\pi k}}  
    e^{\frac{3 \pi \xi_K}{4}    },   
\end{equation}
for $k<{(32/9\pi^2)}\xi_K^{-1}a_\mathrm{end} H_I$, and it is exponentially  suppressed for larger $k$.
Here we have omitted the logarithmic contributions. 
%%%
The exponential suppression for larger $k$ than ${(32/9\pi^2)}\xi_K^{-1}a_\mathrm{end} H_I$ is due to the fact that such a mode has never exited the horizon and does not have time 
to be amplified exponentially. 
This threshold is also confirmed numerically. 
A non-zero $C_{K2}^{{+}}$ means that gauge fields amplified in this mechanism include 
not only positive but also negative frequency modes. 
The asymptotic behavior of the Whittaker function tells that 
the exponential amplification of the gauge fields occurs at the horizon reentry and at a late time 
$k\eta_K \gg 1$, the mode function behaves as
\begin{equation}
A_+(k,\eta)  \sim \frac{1}{2} \sqrt{\frac{\xi_K}{2\pi k}} e^{\pi (\xi_K/2)} \times (\text{oscillation with period } \Delta \eta \simeq k^{-1}),  \quad \text{for} \quad k<{\frac{32}{9 \pi^2}}\frac{a_\mathrm{end} H_I}{\xi_K} \label{asym-mode-kin}
\end{equation} 
up to the phase factor
without any further amplifications.
Here we have used the asymptotic behavior of the Whittaker function, $\lim_{x\rightarrow \infty} W_{\pm i\kappa,\mu}(\pm ix)  = e^{\mp ix/2} (\pm ix)^{\pm i\kappa} = e^{-\pi \kappa/2} e^{\mp i (x/2-\kappa \log[x])}$. 
While a part of the exponential amplifications of the gauge fields during inflation~\cite{Anber:2006xt,Jimenez:2017cdr}, 
\begin{equation}
    A_+(k,\eta \rightarrow 0) \simeq 
    {\frac{1}{\sqrt{\pi \xi_I}}}\exp[\pi \xi_I]/\sqrt{2 k},   
    \label{asym-mode-inf}
\end{equation}  
is explained 
by the asymptotic behavior of the Whittaker function at $k \eta \rightarrow -0$, $ \lim_{x\rightarrow 0} W_{-i\kappa,\mu}(-ix) = 1/\Gamma[1-i\kappa] \sim \exp[\pi \kappa/2]/\sqrt{\pi \kappa}$,
those during kination is explained by the exponentially large coefficients in front of the Whittaker function, 
which overwhelms the exponential suppression of the asymptotic behavior of the Whittaker function at 
$k\eta \rightarrow \infty$. 
At the matching time $\eta=\eta_\mathrm{end}$, the positive and negative frequency modes
are canceled each other so that the gauge fields stay small.

\begin{figure}[t]%%%%%%%%%%%%%%%%%%
\centering
\includegraphics[width=.60\textwidth ]{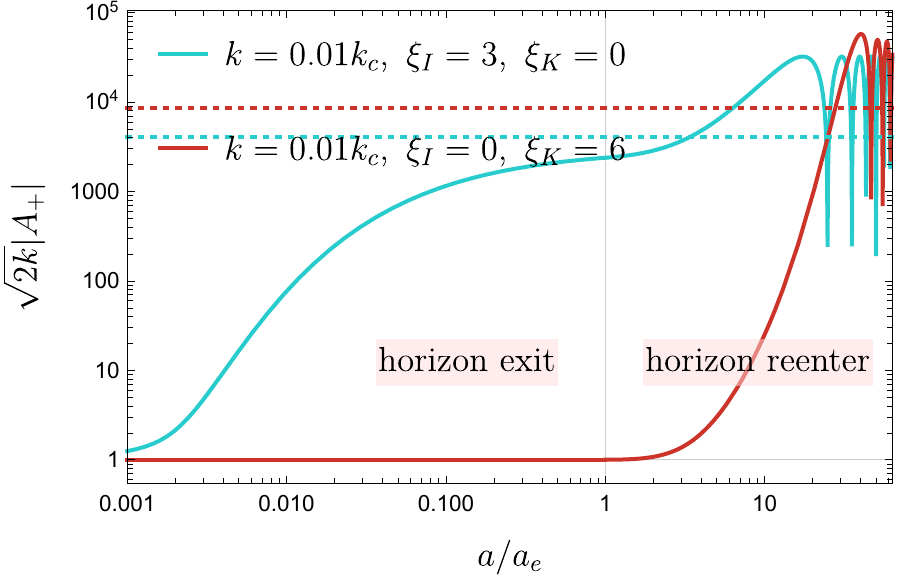}
\caption{
Typical time {evolutions of the gauge field normalized by the momentum, $\sqrt{2k}|A_{I,+}|$, are shown.}
The red and blue lines describe the case of $\xi_I=3,~\xi_K=0$ and $\xi_I=0,~\xi_K=6$, {respectively}.
In the former case, the gauge field is amplified at the horizon exit during inflation, while in the latter case the amplification occurs at horizon reentry during the kination era.
{The red and blue dotted lines represent the approximate formula of gauge field amplification given in Eq.~\eqref{asym-mode-kin} and Eq.~\eqref{asym-mode-inf}, respectively.}
}
\label{fig_field_dynamics}
\end{figure}

To see explicitly the amplification mechanism described in the above, 
in Fig.~\ref{fig_field_dynamics}  we show the typical evolution of the mode function of 
the gauge fields with positive helicity $A_+$ for $(\xi_I, \xi_K) = (0,6)$ as well as (3,0) for comparison with Eqs.~\eqref{solinf} and 
\eqref{eq_generalSolKination} together with the matching condition Eq.~\eqref{eq_determine_C12}. 
Here we take $k=0.01k_c$ with $k_c$ being the horizon scale at the end of inflation, $k_c\equiv a_\mathrm{end}H_I$.
The gauge field amplification during inflation (Eq.~\eqref{asym-mode-inf}) 
can be seen for the case with $\xi_I>0$, 
in which
the gauge field is amplified at horizon exit. 
On the other hand, for the case $\xi_K>0$ the gauge field is amplified at the horizon reentry.
This is the unique characteristics of the magnetogenesis during the kination era.
In realistic models of inflation, $\xi_I$ becomes large at the end of inflation, 
and our treatment of the evolution of $\xi$ (Eq.~\eqref{eq_xis}) does not hold. 
An exponential amplification with $\xi\gg 1$ can take place without the kination era, 
but only for the modes that exit the horizon around the end of inflation.
One may think that
practically the amplification in Eq.~\eqref{asym-mode-kin} is not 
a significant amplification and the late-time observables are not different compared to realistic inflation models (with the instant reheating). 
In this case, however, the exponential amplification $\sim \exp [ \pi \xi_K/2]$ takes place 
not only for the modes that exit the horizon around the inflation end but also 
for the modes that exited the horizon much before the end of inflation and reenter the horizon
during the kination era, 
which shows the significant difference to the instant reheating case.
Indeed, the momentum of the most amplified mode is almost the same, as we will see, 
but the shape of the spectrum is different, 
which we will not explore in depth. 

%%%%%%%%%%%%%%%%%%%%

Now we are ready to evaluate the characteristic properties of the produced gauge fields, 
that is, the
energy density and coherent length.
The physical energy density of the gauge field, 
which can be divided into that of the electric and magnetic fields, 
is estimated as
\begin{align}
    \rho_{EE}(\eta)
	&\equiv
	\frac{1}{2} \braket{\bm{E}^2_\mathrm{p} (\eta)}
	=\frac{1}{2a^4} \int \frac{\df ^3 \bm k }{(2\pi)^3} \left| \p_\eta A_+(k,\eta) \right|^2
	,
	\label{eq_rho_EE}
	\\
	\rho_{BB}(\eta)
	&\equiv
	\frac{1}{2} \braket{\bm{B}^2_\mathrm{p} (\eta)}
	=\frac{1}{2a^4} \int \frac{\df ^3 \bm k }{(2\pi)^3}  \bm k^2\left|  A_+(k,\eta) \right|^2
	.
	\label{eq_rho_BB}
\end{align}
Note that this integral formally suffers from the UV divergence and requires an 
appropriate renormalization~\cite{Ballardini:2019rqh}. Here we shall evaluate them simply by setting the 
UV (and IR) cutoff, which corresponds to the mode that has experienced the tachyonic instability during the kination era in a similar way adopted
in Ref.~\cite{Jimenez:2017cdr}, which is consistent with the results in Ref.~\cite{Ballardini:2019rqh}.
{Namely, we set the UV and IR cutoffs, $k_\mathrm{max}$ and $k_\mathrm{min}$, as
$k_{\rm min} \sim 1/\eta_{K}$ and  $k_{\rm max} =\xi_K/\eta_K^\mathrm{end} = 2\xi_K a_{\rm end} H_I$, respectively,}
with $\eta_K^\mathrm{end} \equiv \eta_\mathrm{end}+{3}/({2 a_\mathrm{end} H_I}) = 1/({2 a_\mathrm{end} H_I})$. 
$k_{\rm mim}$ is the mode which reenters the horizon at $\eta_K$, and 
$k_{\rm max}$  is the mode which 
{experiences the tachyonic instability only at} the end of inflation. 
Note that the mode function has the intrinsic cutoff at $(32/9\pi^2)\xi_K^{-1}a_\mathrm{end} H_I$, 
which is smaller than the UV cutoff for the evaluation of the integral set in the above.
Using  the approximation formula in Eq.~\eqref{asym-mode-kin}, the asymptotic behaviors of $\rho_{EE}$ and $\rho_{BB}$ at later times are analytically evaluated with Eqs.~\eqref{eq_rho_EE} and \eqref{eq_rho_BB} as
\begin{align}
    \rho_{EE} (\eta)
    & =
    \frac{H_I^4}{32 \pi^3 } \left(\frac{a_\mathrm{end} }{ a(\eta)}\right)^4 \frac{e^{\pi \xi_K}}{\xi_K^3} \int_{2\xi_K \eta_K^\mathrm{end}/\eta_K}^{2 \xi_K^2}   \mathrm{d}x  x^3 \left|\frac{\partial_{\eta_K} A_+(\eta_K, x)}{ \sqrt{a_\mathrm{end} H_I x/2 \pi} e^{\pi \xi_K/2}/2}\right|^2 \notag \\
&  \sim H_I^4 \left(\frac{a_\mathrm{end}}{a(\eta)}\right)^4 \frac{e^{\pi \xi_K}}{\xi_K^3}   		, \label{rhoee}
    \\
    \rho_{BB}(\eta)
    & =   \frac{H_I^4}{32 \pi^3 } \left(\frac{a_\mathrm{end} }{ a(\eta)}\right)^4 \frac{e^{\pi \xi_K}}{\xi_K^3} \int_{2\xi_K \eta_K^\mathrm{end}/\eta_K}^{2 \xi_K^2}   \mathrm{d}x  x^3  \left|\frac{A_+(\eta_K, x)}{\xi_K e^{\pi \xi_K/2}/2\sqrt{2 \pi a_\mathrm{end} H_I x}}\right|^2  \notag \\
& \sim H_I^4 \left(\frac{a_\mathrm{end}}{a(\eta)}\right)^4 \frac{e^{\pi \xi_K}}{\xi_K^3},      \label{rhobb}
\end{align}
for $\eta_K >  2 \xi_K \eta_K^\mathrm{end}$, where we have introduced the dimensionless conformal time $x= \xi_K k/(a_\mathrm{end} H_I)$ normalized by the momentum. 
Note that from Eq.~\eqref{asym-mode-kin} we have estimated $\partial_\eta A_+ \simeq k A_+ = (a_\mathrm{end} H_I/\xi_K) x A_+$.
At a sufficiently late time, $\eta_K  >  {(9\pi^2/16)} \xi_K \eta_K^\mathrm{end}$, 
when the lower bound of the integral becomes smaller than the unity, 
the integral becomes a constant and  independent of $\xi_K$ or $H_I$ due to the intrinsic cutoff or the peak of the integrand at $x \simeq {32/9\pi^2}\sim 1$.
This means that
the amplification of the total energy density of the gauge fields terminates
at $\eta_K  \simeq  {(9\pi^2/16)} \xi_K \eta_K^\mathrm{end}$
and dilutes after that according to the cosmic expansion.
Since the gauge fields propagate almost freely after their amplification terminates, 
the energy density of the electric and magnetic fields equilibrate, $\rho_{EE} \simeq \rho_{BB}$, 
during the kination era until
the charged particles start to screen the electric fields. 
We will discuss the screening effect in {Sec.~\ref{sec_schwinger}}.

\begin{figure}[htbp]%%%%%%%%%%%%%%%%%%
\centering
\includegraphics[width=.90\textwidth ]{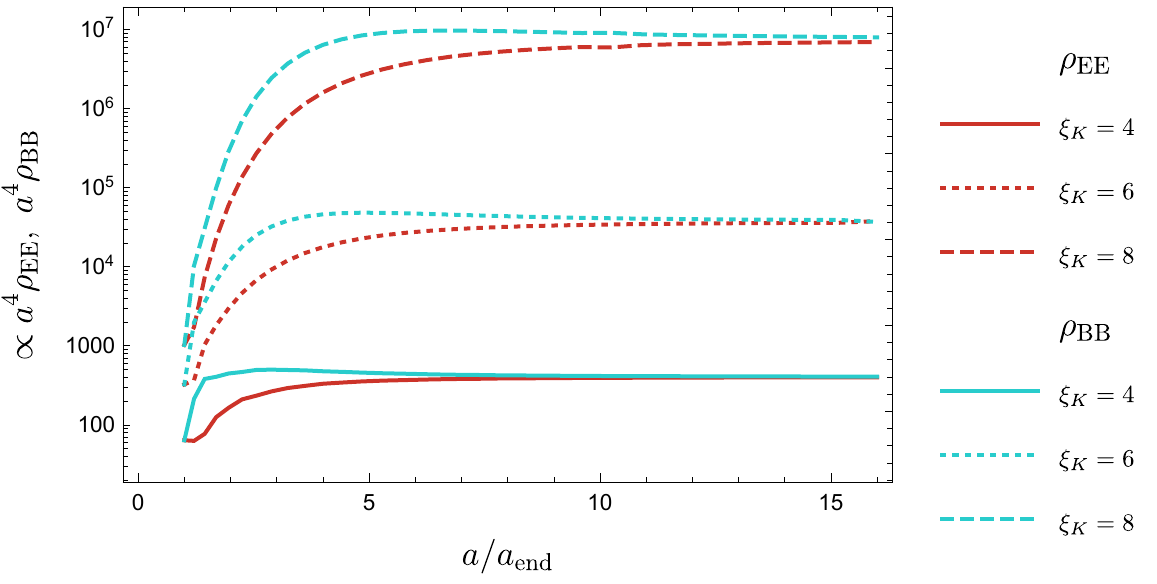}
\caption{
    Typical time evolution of energy density of the gauge fields with $\xi_I=0$ and $\xi_K> 0$ are shown. 
     We numerically evaluate the energy density of the electric and magnetic field in Eqs.~\eqref{eq_rho_EE} and \eqref{eq_rho_BB}.
     The energy densities are given in the comoving quantities, 
     $a^4\rho_{\rm EE}$ and $a^4\rho_{\rm BB}$, for red and blue lines, respectively. 
     Solid, dotted, and dashed lines denote the cases $\xi_K = 4$, 6 and 8, respectively.
}
\label{fig_rhoEEBB_time}
\end{figure}%%%%%%%%%%%%%%%%%%%%

While Eqs.~\eqref{rhoee} and \eqref{rhobb} show the approximate parameter dependence of the 
energy densities of the gauge fields analytically, we need to perform the numerical integration 
to obtain the quantitative estimates.
By performing the integration in Eqs.~\eqref{rhoee} and \eqref{rhobb}, 
we obtained $\rho_{EE}$ and $\rho_{BB}$ as the function of the scale factor $a$ after inflation.
We show the typical gauge field amplifications with $\xi_K=4,~6$, and $8$ in Fig.~\ref{fig_rhoEEBB_time}.
We can see that the gauge field amplification in the comoving values 
is saturated around 
\begin{align}
a = a_{\rm sat} \simeq \frac{3\pi \sqrt{\xi_K} }{4} a_\mathrm{end},
\end{align}
and becomes constant after that, which is consistent with the estimates 
Eqs.~\eqref{rhoee} and \eqref{rhobb}.
With these numerical calculations, quantitatively we find
\begin{align}
\rho_{EE}(\eta)
		&=
			{1}
			\times 10^{-2}
			H_I^4 \left( \frac{a_{\rm  end}}{a(\eta) }\right)^4
			\frac{e^{ \pi\xi_K} }{ \xi_K^3},\label{eqrhoEE}
\\
\rho_{BB}(\eta)
&
=
			1\times 10^{-2}
			H_I^4 \left( \frac{a_{\rm  end}}{a(\eta) }\right)^4
			\frac{e^{ \pi\xi_K} }{ \xi_K^3}, \label{eqrhoBB}
\end{align}
{which are the main results of the present paper.}
In the following we use these fitting formula for the investigation of the consequence of this mechanism.

Some more comments follows. We can see that
the magnetic field is amplified at first, and the electric field catches up it.
This is because in the gauge field amplification during the kination era, 
the gauge fields are amplified when they turn from the superhorizon mode to the 
subhorizon mode and hence the magnetic fields, which are the spatial derivative of the vector field, 
grows faster than the electric fields, which are the time derivative of the vector field.
This is the opposite behavior to the gauge field amplification during inflation, where the gauge fields are amplified when they {\it exit} the horizon. 
In this case, the electric fields
are amplified faster and stay larger than magnetic fields in the instant reheating approximation.
While in the inflation with the instant reheating case, there will be no free propagation of the gauge 
fields and there are no equilibration between the electric and magnetic fields, 
in the kination case,
after the amplification ends and the {relevant} modes enter subhorizon, the gauge fields freely oscillate and  end up with equal energy densities between the electric and magnetic fields.  
See Ref.~\cite{Kobayashi:2019uqs} for a similar discussion on the equilibration of the electric and magnetic fields 
during the reheating era.

The coherence length of the gauge fields is defined by
\begin{align}
    \lambda_{\rm phys}(\eta_K)
    &\equiv
    \frac{1}{\rho_{BB}} \frac{1}{2a^4} 
    \int \frac{\df ^3 \bm k }{(2\pi)^3}   \frac{2\pi a(\eta)}{k} \bm k^2\left|  A_+(k,\eta_K) \right|^2 \notag 
    \\&= 
    \frac{2 \pi \xi_K}{H_I} \left(\frac{a(\eta)}{a_\mathrm{end}}\right)
     	\frac{\int_{{2 \xi_K \eta_K^\mathrm{end}/\eta_K}}^{{2 \xi_K^2}} 	   x^2\df  x  \left| A_+({x},\eta_K) /(\xi_K e^{\pi \xi_K/2}/2\sqrt{2 \pi a_\mathrm{end} H_I x})\right|^2} {\int_{{2 \xi_K \eta_K^\mathrm{end}/\eta_K}}^{{2 \xi_K^2}} x^3\df  x  \left| A_+({x},\eta_K)/(\xi_K e^{\pi \xi_K/2}/2\sqrt{2 \pi a_\mathrm{end} H_I x}) \right|^2}. 
\end{align}
{Once more, with the numerical calculation we find the quantitative fitting formula for the coherence length as}
\begin{align}
		\lambda_{\rm phys}({\eta})
&= 
			 {0.13} 
			 \frac{2\pi }{  H_I} 
			 \left( \frac{a({\eta})}{a_{\rm  end}} \right)\xi_K, 
	    \label{eq_lambda_num}
\end{align}
which is similar to the one obtained in the gauge field amplification at the end of inflation 
with the instant reheating approximation, $\lambda_\mathrm{phys} \simeq 0.6 (2 \pi/H)\xi_I $~\cite{Jimenez:2017cdr}.

For later purpose we also {evaluate} the comoving helicity density, 
\begin{align}
	h_{c}
	&\equiv
	a^2\braket{\bm{A}\cdot\bm{B}_\mathrm{p}}
	=
	 \int \frac{\df ^3 \bm k }{(2\pi)^3} k \left|  A^k_+ \right|^2
	=
	{a^3 } \frac{\rho_{\rm BB}\lambda_{\rm phys}}{\pi}
	,
	\label{eq_h_c}
\end{align}
which is quantitatively estimated by using Eqs.~\eqref{eqrhoBB} and \eqref{eq_lambda_num} as
\begin{align}
		h_{\rm c}({\eta})
&= 
		{3\times 10^{-3} }a_\mathrm{end}^3 
		H_I^3  
		\frac{e^{ \pi\xi_K} }{ \xi_K^2}
		\label{eq_hc_num}
.\end{align}
Since only the plus mode is amplified, the gauge fields are maximally helical 
and the electric and magnetic fields effectively runs parallel.
{Note that the helicity density is related to}
the cross correlation of $E_\mathrm{p}$ and $B_\mathrm{p}$, {$\rho_{EB} \equiv  \braket{\bm{E}_\mathrm{p}\cdot \bm{B}_\mathrm{p}}$,}
as 
\begin{align}
	\frac{\df}{\df \eta } h_c = -2 a^4 \braket{\boldsymbol E_\mathrm{p}\cdot \boldsymbol B_\mathrm{p}}
	= -2 a^4\rho_{\rm EB}.
	\label{eq_rhoeb_hc}
\end{align}
{Comparing with $\rho_{EE}$ and $\rho_{BB}$, we roughly estimate $\rho_{EB}$ as
\begin{equation}
\rho_{EB} \simeq 1\times 10^{-2} H_I^4 \left(\frac{a_\mathrm{end}}{a(\eta)}\right)^4 \frac{e^{\pi \xi_K}}{\xi_K^3}, 
\end{equation}
which is used to evaluate the backreaction on the ALP dynamics from the gauge field amplification
in the next subsection.}

%%%%%%%%%%%%%%%%%%%%%%%%%%%%%%%%%%%%%%%%%%%%
\subsection{Constraints from backreaction}
\label{sec_gaugefieldproduction_backreaction}
%%%%%%%%%%%%%%%%%%%%%%%%%%%%%%%%%%%%%%%%%%%%

The explosive amplification of gauge fields {causes} the backreaction on the {cosmic expansion and} the ALP dynamics Eq.~\eqref{eq_eom_alp}. 
{If the backreaction is too large, the investigation in the previous subsection is spoiled.}
For the inflation and the kination era, we {define} the {following} two {quantities that describe} the back reaction, {namely}, (1) the ratio of the energy density of the hypergauge fields to that of the ALP on the Friedman equation and (2) the ratio of the source term to the Hubble friction term on the Klein-Gordon equation Eq.~\eqref{eq_eom_alp}, represented $\delta_F$ and $\delta_K$, respectively, 
{as} \cite{Jimenez:2017cdr}
\begin{align}
	\delta_F 
	\equiv
	\frac{\rho_{EE}+\rho_{BB}}{3H^2 \Mpl^2}
\quad ,\quad
	\delta_K 
	\equiv
	\left|
	\frac{\rho_{EB}/\Lambda}{3H \dot\phi} 
	\right|  
    =
    \frac{\xi_K}{3}
	\frac{|\rho_{EB}|}{H^2 \Mpl^2} 
	\label{eq_backreacts}
	,
\end{align}
{If $\delta_F\gg 1$, the assumption that the Universe is dominated by the inflaton is broken down. 
Note that it may corresponds to a realization of reheating if it occurs during kination. 
However, in order to see if the Universe really becomes the thermal radiation dominated, 
one needs to investigate the particle production from the gauge fields, or the Schwinger effect,
which we will discuss it in the next section. In such a case, it is difficult to perform a consistent
analysis. 
If $\delta_K\gg1$, additional friction term on the ALP dynamics overwhelms the Hubble friction, 
which would make the instability parameter $\xi$ much smaller 
and hence the gauge amplification is suppressed.
For the consistency of our analysis, therefore,} 
we require $\delta_F,~\delta_K<1$ until the {saturation} of gauge field amplification.

Let us first consider the backreaction on the cosmic expansion $\delta_F$. 
Since we have estimated the gauge field amplification saturates at $a_\mathrm{sat} $ 
without backreaction, we require $\delta_F<1$ at $a = a_\mathrm{sat}$.  
It is evaluated as
\begin{equation}
    \delta_F(a_\mathrm{sat} )  \simeq \frac{2}{3} \times 10^{-2} \left(\frac{H_I}{\Mpl}\right)^2 \left(\frac{a_\mathrm{sat}}{a_\mathrm{end}}\right)^2 \frac{e^{\pi \xi_K}}{\xi_K^3} 
    \simeq 
    3.7 \times 10^{-2}  \left(\frac{H_I}{\Mpl}\right)^2  \frac{e^{\pi \xi_K}}{\xi_K^2}. 
\end{equation}
For $H_I \simeq 10^{13}$ GeV, $\delta_F<1$ is satisfied for $\xi_K<10$. 
On the backreaction on the ALP dynamics, $\delta_K$, at the saturation of the gauge field amplification, it is evaluated as
\begin{equation}
    \delta_K(a_\mathrm{sat}) 
    \simeq  
    \frac{\xi_K}{3} \times 10^{-2} \left(\frac{H_I}{\Mpl}\right)^2 \left(\frac{a_\mathrm{sat}}{a_\mathrm{end}}\right)^2 \frac{e^{\pi \xi_K}}{\xi_K^3} 
    \simeq 
    1.9 \times 10^{-2}  \left(\frac{H_I}{\Mpl}\right)^2  \frac{e^{\pi \xi_K}}{\xi_K}, 
\end{equation}
where we have used Eq.~\eqref{eq_xis}. 
For $H_I\simeq 10^{13}$ GeV, the condition $\delta_K(a_\mathrm{sat}) <1$ 
is satisfied for $\xi_K \lesssim 10$. 

\ 

The discussion in this section does not limited to the SM U(1) gauge fields 
but is also applicable to the any dark U(1) gauge fields.
In the next section, we will investigate phenomena inherent in the SM gauge field 
or in the presence of the charged particles, namely, the Schwinger effect.

%%%%%%%%%%%%%%%%%%%%%%%%%%%%%%%%%%%%%%%%%%%%
\section{Schwinger effect during kination and reheating}
\label{sec_schwinger}
%%%%%%%%%%%%%%%%%%%%%%%%%%%%%%%%%%%%%%%%%%%%

Thus far, we have studied the dynamics of the system only with the runaway ALP and the hypergauge field. 
Once we take into account the matter field in the system, there arises an inevitable effect on the dynamics of the system.
Namely, the amplified hypergauge fields induce pair production of particle and antiparticle charged under the hypergauge interaction~\cite{Heisenberg:1935qt,Schwinger:1951nm}, which is known as the Schwinger effect.
This effect hinders the amplification of the gauge field due to the following two reasons: (1) The pair production of charged particles acts as the friction term for the gauge field amplification, and (2) the produced charged particles screen the (hyper) electric field.
In particular, the produced charged particles are eventually thermalized, 
which opens up the possibility to complete the reheating of the Universe 
in the scenario with the kincation era, which is also dubbed as ``Schwinger reheating''~\cite{Tangarife:2017rgl}.
In this section, we discuss these two consequences of the Schwinger effect. 
Indeed, the evaluation of the Schwinger effect with the dynamical gauge field background is quite involved, and to our best of knowledge even a method to analyze the system consistently has not been established. 
Therefore we adopt several simplification and assumptions, mainly following the analysis in Ref.~\cite{Domcke:2018eki}, to get a rough estimate for the consequence of the Schwinger effect and to show the possible realization of the Schwinger reheating\footnote{See also recent study
on the Schwinger effect in axion inflation with the gradient expansion formalism~\cite{Gorbar:2021rlt}, which however requires model-dependent numerical analysis.}.

Hereafter we assume that the Higgs field acquires a sufficiently large induced mass through an appropriate non-minimal coupling to gravity or spectator fields, which keeps the electroweak symmetry unbroken during the period of interest and suppresses the production of the Higgs fields, and focus on the production of the (massless hyper U(1) charged) fermions.
This assumption has also an advantage to avoid the Higgs vacuum instability 
into the unwanted true AdS vacuum.

%%%%%%%%%%%%%%%%%%%%%%%%
\subsection{Review of the Schwinger effect and its application to gauge field amplification during kination}
%%%%%%%%%%%%%%%%%%%%%%%%

We first briefly review the structure of the fermion spectrum in the presence of the gauge field with introducing the Landau levels and how we can estimate the number density of the fermions pair produced by the Schwinger effect.
Let us consider a massless Dirac fermion $\psi$ with a hypercharge $Q$ with the Lagrangian Eq.~\eqref{eq_lagrangian_fermi}.
The equation of motion for the fermion in the presence of a ``background'' gauge field $A_\mu$ is given as
\begin{align}
    \left[ a^{-1} \gamma^\mu \left( \partial_\mu +ig'Q A_\mu \right) + \frac32 H\gamma^0 \right] \psi = 0. \label{prepsieom}
\end{align}
Since massless fermions and gauge fields are conformal, we can eliminate $a$ and $H$ by rescaling $\tilde{\psi} \equiv a^{3/2}\psi, \tilde{A}_\mu \equiv A_\mu, \tilde{A}^\mu \equiv a^2 A^\mu$.  Eq.~\eqref{prepsieom} is then rewritten as 
\begin{align}
    \gamma^\mu \left( \partial_\mu +ig'Q \tilde{A}_\mu \right) \tilde{\psi} = 0. \label{psieom1}
\end{align}

It is desirable if we could solve the quantum evolution equation for both the gauge fields and the fermions simultaneously, but it is technically difficult.
Instead, we take the gauge field amplified by the ALP dynamics as the background field and examine the spectrum and dynamics of the fermions with the consistency conditions. 
To make the analytic estimate possible, we employ an approximation that the gauge fields are uniformly distributed while maximally helical to grasp the properties of the gauge field of our interest.
Namely, we write the background gauge field as $\tilde{A}_\mu = (0, 0, B_c x, -E_c \eta)$ ($z$-axis is the direction of the electromagnetic field.) so that the electric and magnetic fields are parallel as discussed in Sec.~\ref{sec_gaugefieldproduction}. 
This approximation corresponds to a treatment that we stop the gauge field amplification at a certain time and take one patch within its correlation length.
We also ignore the cosmic expansion and substitute $a = 1$, which means that we replace $\eta$ with $t$. 
This also suggests that the hyper electric and magnetic fields we discuss here are the {\it comoving} ones, but not the {\it physical} ones discussed in the previous section.
We added the subscript $c$ to indicate that explicitly.
These assumptions are justified if the following two conditions are satisfied. 
First, the time scale of the fermion production is not much slower than that of the gauge field time evolution and the Hubble time scale.
Second, the gauge field coherence length and the Hubble length is much larger than the one that corresponds to the typical energy scale of the produced fermions.
We will take the time scale of the fermion production as the Hubble time at $a = a_{\rm sat}$, which is the same order as the time scale of the gauge field evolution (see Fig.~\ref{fig_rhoEEBB_time}), so that the former condition is satisfied.
We will confirm the latter condition later.

Let us now investigate the Schwinger mechanism. 
Eq.~\eqref{psieom1} is rewritten as
\begin{align}
    \left[ \partial_t + s\bm{\nabla} \cdot \bm{\sigma} - isg'Q(B_c x\sigma_y - E_c t\sigma_z) \right] \tilde{\psi}_{L/R} =0, \label{preweyleom}
\end{align}
where the Dirac fermion is decomposed into the chiral (Weyl) components $\tilde{\psi}_{L/R}$. $s$ takes $+1$ and $-1$ for left- and right-handed component, respectively. 
By introducing an auxiliary field $\Psi_{L/R}$ as
\begin{align}
    \tilde{\psi}_{L/R} \equiv \left[ -\partial_t + s\bm{\nabla} \cdot \bm{\sigma} - isg'Q(B_c x\sigma_y - E_c t\sigma_z) \right] \Psi_{L/R},
\end{align}
Eq.~\eqref{preweyleom} becomes
\begin{align}
    \left[ -\partial_t^2 + \bm{\nabla}^2 - 2ig'Q(B_c x\partial_y - E_c t\partial_z) - g'^2 Q^2 (B_c ^2 x^2 + E_c ^2 t^2) + g'Q(B_c  + isE_c )\sigma_z \right] \Psi_{L/R} = 0. \label{prechiraleom}
\end{align}
By performing the (partial) Fourier transformation of $\Psi_{L/R}$ with respect to the spatial coordinates $y$ and $z$ as
\begin{align}
    \Psi_{L/R}(t,\bm{x}) = \int\frac{\df k_y \df k_z}{(2\pi)^2} e^{i(k_y y + k_z z)} \Psi_{L/R}(t, x, k_y, k_z),
\end{align}
Eq.~\eqref{prechiraleom} is rewritten as
\begin{align}
    \left[ -\partial_t^2 + \partial_x^2 - (g'QB_c x - k_y)^2 - (g'QE_c t + k_z)^2 + g'Q(B_c  + isE_c )\sigma_z \right] \Psi_{L/R} = 0. \label{eqmixed}
\end{align}
By redefining the coordinate as $X \equiv \sqrt{g'|Q|B_c}x - \varsigma k_y/\sqrt{g'|Q|B_c}$, where $\varsigma$ is the sign of $Q$, Eq.~\eqref{eqmixed} turns into the form that can be solved by the method of the separation of variables with decomposing the field as $\Psi_{L/R}(t, x, k_y, k_z) \equiv h_n(X) g_{L/R}(t, n, k_z)\chi_\varsigma$, where $\sigma_z \chi_\varsigma = \varsigma\chi_\varsigma$. 
It is clear that the $X$-dependent part of the equation of motion, 
$(\partial_X^2 -X^2) h_n(X)= -(2n+1) h_n(X)$, is the same as that of the harmonic oscillator and it has a solution with discretized energy levels labeled by a non-negative integer $n = 0, 1, 2, \hdots$, as
\begin{align}
    h_n(X) = \frac{1}{\sqrt{2^n n!}} \left(\frac{g'|Q|B_c}{\pi}\right)^{1/4} e^{-X^2/2} H_n(X),
\end{align}
where $H_n(X)$ is the Hermite polynomial. 
Here we have normalized the $X$-dependent part so that it satisfies 
$\int dx |h_n(X)|^2 =1$.
The equation of motion for the $X$-independent part reads
\begin{align}
    \left[ \partial_t^2 + (g'QE_c t + k_z)^2 - g'|Q|(B_c + isE_c) \right] g_{L/R} = -(2n+1) g'|Q|B_c g_{L/R}. \label{eomglr}
\end{align}
Note that $X$-independent part $g_{L/R}$ depends on the energy level $n$ but not $k_y$ since $X$ already contains $k_y$. 

To see the spectrum of the system, we first clarify the case with vanishing electric field $E_c = 0$.
The solution of Eq.~\eqref{eomglr} is then just a plane wave, 
\begin{align}
    g_{L/R} = \frac{1}{\sqrt{2\omega_{L/R}}} e^{i \omega_{L/R}t}
\end{align}
with the dispersion relation
\begin{align}
    \omega_{L/R} = \begin{cases}
      \pm \sqrt{k_z^2 + 2ng'|Q|B_c} & (n = 1, 2, \hdots) \\
      -s\varsigma k_z & (n = 0)
    \end{cases}. \label{disp_rel}
\end{align}
This discrete energy level is nothing but the relativistic Landau level. 
The discretization of the energy levels can be understood by the consequence that the uniform magnetic fields restrict the transverse motion of the charged particles by the Lorentz force.
We can see that the $n=0$ level called the lowest Landau level (LLL) has a unique dispersion relation where the negative frequency is continuously connected to the positive frequency in an opposite way for left- and right-handed fermions (with the same sign of charge). 
This is because the LLL is understood as a fermion moving in the direction of the magnetic field with aligning its spin (anti)parallel to the same direction. 
On the other hand, the $n \geq 1$ levels called the higher Landau levels (HLL) have the same structure for the right- and left-handed fermions. 
As a result, while the HLL contribute to the particle production without chirality, the LLL contribute to the chiral charge of the pair-produced particles.

Let us now examine the particle production when we apply the electric field
in the system where the Landau levels are formed. 
Since the states are not well-defined under a non-vanishing electric field due to the explicit time dependence of the action, we consider the case where we apply a constant electric field parallel to the magnetic field as described in the above only for a certain time duration $0 < t < \tau_{\rm prod}$, with the initial condition where the system is in the vacuum state so that the fermion states are filled up to $\omega_{L/R}=0$. 
For the LLL, the fermions are accelerated up to $k_z = g' Q E_c \tau_{\rm prod}$ along the dispersion relation for the LLL, which means that the pair-created particles are a right-handed fermion and a left-handed antifermion for $Q>0$. Similar arguments apply for $Q<0$.
The (comoving) number density of the produced particles at the LLL is then evaluated as \cite{Domcke:2018eki}
\begin{align}
    n_\psi^{\rm LLL} &= 2\times \frac{1}{V} \int \df^3 x \int \frac{\df k_y \df k_z}{(2\pi)^2} [h_0(X)]^2 \Theta(-k_z)
    \Theta(k_z + g'|Q|E_c\tau_{\rm prod}) 
    \nonumber \\
    &= \frac{g'^2 |Q|^2}{4\pi^2} E_c B_c \tau_\mathrm{prod},  \label{nLLL} 
\end{align} 
where $V$ denotes the volume of the system. 
The prefactor 2 counts the right-handed fermion and the left-handed antifermion.
One can see that this process is consistent with the chiral anomaly. 
This is not just a coincidence but the consequence of the chiral anomaly as is seen in the discussion of Nielsen and Ninomiya~\cite{Nielsen:1983rb}.
On the other hand, for the HLL, due to the nonzero electric field, the positive and the negative frequency modes for given $n$ are mixed and the fermion and the antifermion for both left and right-handed particles are produced, which is understood as the quantum tunneling process between the energy gap. 
Then the (comoving) number density of the produced particles is evaluated as
\begin{align} 
    n_\psi^{(n)} &= 4\times \frac{1}{V} \int \df^3 x \int \frac{\df k_y \df k_z}{(2\pi)^2} [h_0(X)]^2 \Theta(-k_z) \Theta(k_z + g'|Q|E_c \tau_{\rm prod}) e^{-2\pi n B_c/E_c} \nonumber \\
    &= \frac{g'^2 |Q|^2}{2\pi^2} E_c B_c \tau_\mathrm{prod} e^{-2\pi n B_c/E_c}. \label{nHLL}
\end{align}
The prefactor 4 counts both the particle and the antiparticle of the left and right-handed fermions.

Before proceeding, let us clarify how we shall apply these results to the case of our interest. 
When the hypergauge fields are amplified during kination, we have seen that their amplification is saturated at $a = a_{\rm sat} \sim \frac{3\pi}{4} \sqrt{\xi_{\rm eff}} a_\mathrm{end} \gg a_\mathrm{end}$ (see Fig.~\ref{fig_rhoEEBB_time}). 
With focusing on the last minute of the hypergauge field amplification, we take $\tau_{\rm prod}$ as the Hubble time at $a = a_{\rm sat}$, or $\eta_\mathrm{prod}= 1/(2a_{\rm sat}H_{\rm sat})$ (see Eq.~\eqref{eq_etakin}),  for the evaluation of the properties of the pair-produced particles.
As we have mentioned, this treatment justifies our approximation that the electric and magnetic fields are taken to be a constant in time. 
The (comoving) electric and magnetic fields can be expressed in terms of the physical ones as $E_c = a^2 E_\mathrm{p}$ and $B_c = a^2 B_\mathrm{p}$.
In the case where there are multiple (Weyl) fermions in the system as in the SM, we can just replace $|Q|^2$ in Eqs.~\eqref{nLLL} and \eqref{nHLL} with an effective value, $Q_2{/2} \equiv \sum_i |q_i|^2{/2}$, with $q_i$ being the hypercharge of each (Weyl) fermion, $|Q|^2 \to Q_2/2$.
Note that it is not the square of the sum of charges of each particle $\left( \sum_i |q_i| \right)^2$,
and the factor 1/2 is multiplied due to the difference between the Weyl and Dirac representation. 
(For later convenience, we introduce the notation $Q_n \equiv \sum_i |q_i|^n$ and distinguish it from $Q_1^n = (\sum_i |q_i|)^n$.)
In summary, with Eqs.~\eqref{nLLL} and \eqref{nHLL} (divided by the factor of $a_\mathrm{sat}^3$), the physical number density of the LLL as well as HLL at the time of the saturation of the gauge field amplification is given by
\begin{align}
    n_\psi^{\rm LLL} (a_\mathrm{sat}) &= \frac{g'^2 Q_2}{16\pi^2} \frac{E_\mathrm{p} B_\mathrm{p}}{H_{\rm sat}},  \label{nLLLphys} \\
    n_\psi^{(n)}(a_\mathrm{sat}) &= \frac{g'^2 Q_2}{8\pi^2} \frac{E_\mathrm{p} B_\mathrm{p}}{H_{\rm sat}} e^{-2\pi n B_\mathrm{p}/E_\mathrm{p}}. \label{nHLLphys}
\end{align}  

Furthermore, we assume that the Schwinger pair production after the saturation of the hypergauge field amplification is negligibly small since the hypergauge field starts to oscillate and they are no longer constant, which violates our assumption for the particle production.
The number density of the particles (Eqs.~\eqref{nLLL} and \eqref{nHLL}) is assumed to be redshifted according to the cosmic expansion after that. 
In this point of view, we give a conservative estimate of the number density of the produced particles.

%%%%%%%%%%%%%%%%%%%%%%%%%%%%%%%%%%%%%
\subsection{Backreaction on gauge field amplification} \label{subsec_BRgauge}
%%%%%%%%%%%%%%%%%%%%%%%%%%%%%%%%%%%%%

Once the Schwinger effect becomes sufficiently effective, induced (hyper) electric current from the pair-produced charged particles is no longer negligible for the amplification of the hypergauge field.
Once more, it is desirable if we could solve the evolution of the system simultaneously, but the equations of motion are highly non-linear and too difficult to solve with the best of our knowledge and technique.
Instead, here we take into account the backreaction of the Schwinger effect on the gauge field dynamics by giving some assumptions, following Ref.~\cite{Domcke:2018eki}.

The equation of motion for the physical energy density of the gauge fields in the presence of the background ALP dynamics and the induced current reads
\begin{equation}
    \frac{d}{d t}(\rho_{EE}+ \rho_{BB})=-4 H (\rho_{EE}+ \rho_{BB}) + 2 \xi_K H \langle {\bm E}_\mathrm{p} \cdot {\bm B}_\mathrm{p}\rangle - \left\langle {\bm E}_\mathrm{p} \cdot g' \sum_i q_i  {\bm J}_i\right\rangle,   \label{eomende}
\end{equation}
where ${\bm J}_i$ is the induced matter current of the particle $i$.
By using the mode functions in the way described in the above, one can evaluate the induced current $\langle J^{\alpha}_i \rangle \equiv \langle {\bar \psi}_i \gamma^{\alpha} \psi_i \rangle$ in the presence of the homogeneous hyper electric and magnetic fields in $z$ direction as~\cite{Domcke:2018eki}
\begin{equation}
    \frac{1}{a^3} g' \sum_i q_i \langle J_i^{z} \rangle
    \simeq   \frac{g'^3 Q_3}{4 \pi^2} {\coth\left(\frac{\pi B_\mathrm{p}}{E_\mathrm{p}}\right)} \frac{E_\mathrm{p}B_\mathrm{p}}{H}. \label{indcurrent}
\end{equation}
We will confirm that the scattering and thermalization is not effective at least during the gauge field amplification and the estimate for the induced current in the above is valid. 
By taking $ \langle {\bm E}_\mathrm{p} \cdot {\bm B_\mathrm{p}}\rangle  = E_\mathrm{p}B_\mathrm{p}$ and $\left\langle {\bm E}_\mathrm{p} \cdot g' \sum_i q_i  {\bm J}_i\right\rangle = g' \sum q_i E_\mathrm{p} \langle J_i^z\rangle$, we can see that the induced current from the Schwinger effect is formally removed in the evolution equation for the energy density of the hypergauge fields~\eqref{eomende} by replacing $\xi_K$ with $\xi_\mathrm{eff}$, which is defined as
\begin{align}
    \xi_{\rm eff} = \xi_K -
     \frac{g'^3 Q_3}{8 \pi^2} \coth\left(\frac{\pi B_\mathrm{p}}{E_\mathrm{p}}\right)  \frac{E_\mathrm{p}}{H^2}.  \label{EMBR2}
\end{align}
Thus once we specify the ``background'' (hyper) electric and magnetic field, we can take into account the backreaction on the gauge field amplification from the Schwinger effect in terms of the effective instability parameter $\xi_\mathrm{eff}$.

One might think that one can just solve the mode equation with $\xi_\mathrm{eff}$. 
However, we would like to take the electric and magnetic fields amplified by the ALP dynamics themselves as the ``background'' fields that cause the Schwinger effect, which appear in the effective instability parameter $\xi_\mathrm{eff}$. 
Thus one cannot solve the mode equation consistently as if $\xi_\mathrm{eff}$ is a constant, and hence further approximation is needed to determine the resultant hyper electric and magnetic field strength with taking into account the backreaction from the Schwinger effect. 
In Ref.~\cite{Domcke:2018eki}, two ways of approximation were proposed. 
The one is to take both side of Eq.~\eqref{eomende} to be zero so that gauge field amplification and the dilution due to the cosmic expansion is equilibrated. 
From the right-hand side, we can evaluate the relationship between the electric and magnetic fields for this equilibrium solution and draw a contour in the $E$-$B$ plane, on which the electric and magnetic fields are ``stable''. 
By looking at this contour, we can determine the upper bound of the magnetic field strength, which is identified as the ``maximal'' solution. 
The other way is to take the expression of the electric and magnetic fields without the Schwinger effect, namely Eqs.~\eqref{eqrhoEE} and \eqref{eqrhoBB}, with replacing $\xi_K$ with $\xi_\mathrm{eff}$ as
\begin{align}
    E_{\rm eff} (\xi_\mathrm{eff}) \approx B_{\rm eff} (\xi_\mathrm{eff}) = \sqrt{2\rho_{\rm BB}} = 0.14 H_I^2 \mathcal{I}(\xi_{\rm eff}) \left(\frac{a}{a_{\rm end}}\right)^{-2}, \label{EMBR1}
\end{align}
where $\mathcal{I}(\xi_{\rm eff}) \equiv e^{\pi\xi_{\rm eff}/2}/\xi_{\rm eff}^{3/2}$, and substitute Eq.~\eqref{EMBR1} into the expression of $\xi_\mathrm{eff}$ in Eq.~\eqref{EMBR2}.
By requiring the resultant $\xi_\mathrm{eff}$ in the same as the input $\xi_\mathrm{eff}$ 
at the time of the saturation of the gauge field amplification, $H=H_\mathrm{sat}$, 
\begin{align}
    \xi_{\rm eff} = \xi_K -
    \frac{g'^3 Q_3}{8 \pi^2} \cot \left(\frac{\pi B_\mathrm{eff}}{E_\mathrm{eff}}\right)   \frac{E_\mathrm{eff}}{H_\mathrm{sat}^2}, \label{EMBR3}
\end{align}
we can determine the self-consistent solution of $\xi_\mathrm{eff}$ as a function of $\xi_K$, which is identified as the ``equilibrium'' solution.

\begin{figure}[htbp]
    \centering
    \includegraphics[width=.60\textwidth]{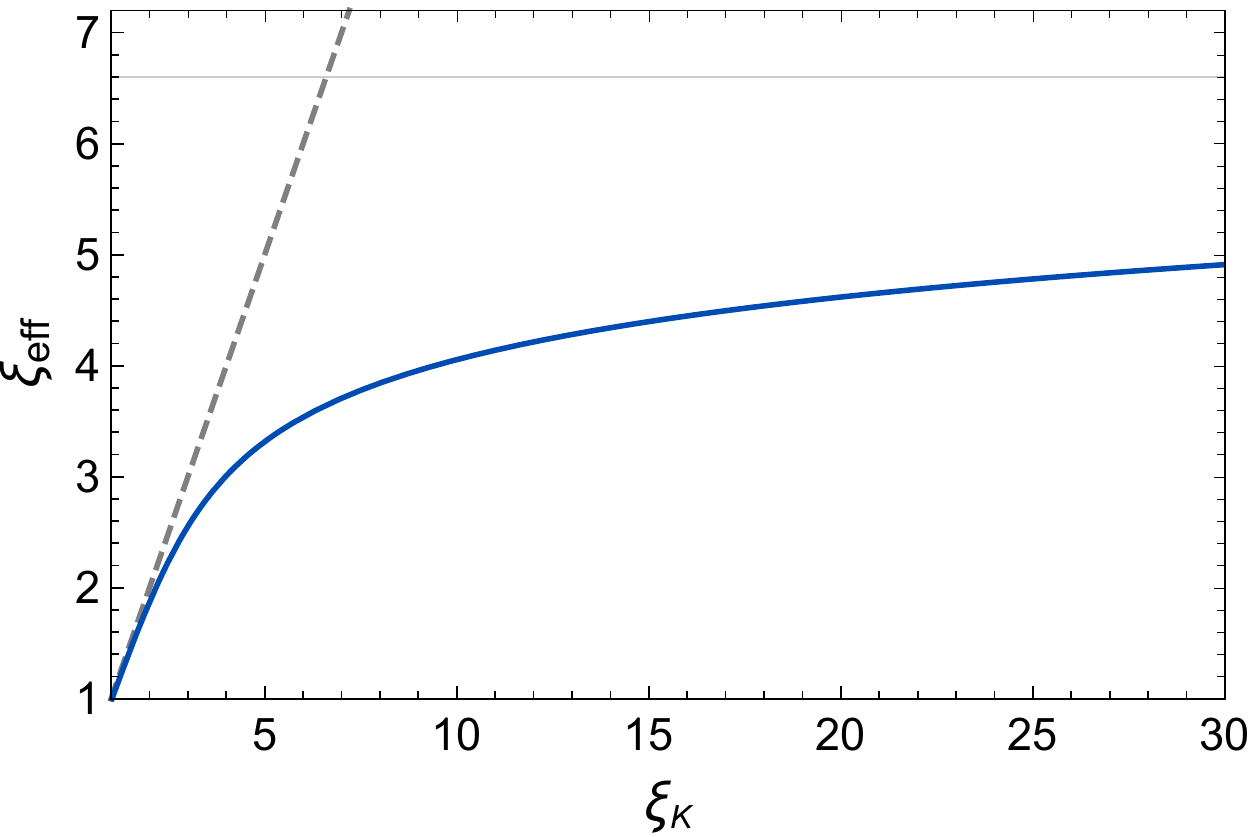}
    \caption{\label{fig:EMBR} ``Equilibrium'' solution of the effective instability parameter $\xi_\mathrm{eff}$ for the hypergauge field amplification as a function of $\xi_K$ in Eq.~\eqref{EMBR2}. Here we take $Q_3 = 41/12, g' = 0.3$ (motivated by the standard model) and $a = a_{\rm sat} = \frac{3\pi}{4} \sqrt{\xi_\mathrm{eff}} a_{\rm end}$. The dashed and the solid lines show the bare value $\xi_K$ and effective value $\xi_{\rm eff}$, respectively. $\xi_{\rm eff}$ is highly suppressed for $\xi_K > 2$ due to the backreaction of the induced current. Here, the horizontal line denotes $\xi_{\rm eff} = 6.6$, below which the pair-produced particles are thermalized before they dominates the Universe for $H_I=10^{13}$ GeV as discussed later in this section.}
\end{figure}

The former solution is reasonable for the inflationary magnetogenesis, since the amplified magnetic fields are expected to be saturated and become constant. 
In the case of kination, which of our interest, however, the Hubble parameter decreases with time and the electric and magnetic fields are not expected to become a constant when the gauge field amplification saturates, as we have seen in the case without Schwinger effect Eqs.~\eqref{eqrhoEE} and \eqref{eqrhoBB}. 
Therefore, we take the latter, ``equilibrium'' solution, to be our rough estimate of the electric and magnetic field strength from the ALP dynamics during kination in the presence of the Schwinger effect.
Here we evaluate the parameters at the time of gauge field saturation at $a=a_\mathrm{sat}$.
The numerical result of the ``equilibrium'' solution for the effective instability parameter $\xi_\mathrm{eff}$ is depicted in Fig.~\ref{fig:EMBR}, where we have taken $Q_3 = 41/12  \simeq 3.4$, which is motivated from the particle contents in the standard model, and $g'=0.3$ as well as $a_\mathrm{sat} = \frac{3\pi}{4} \sqrt{\xi_\mathrm{eff}} a_{\rm end}$. 
We can see that while $\xi_\mathrm{eff} \simeq \xi_K$ for $\xi_K \lesssim 2$, it is highly suppressed for $\xi_K > 2$ due to the backreaction from the Schwinger effect and it grows only logarithmically. 
The typical energy carried by the LLL fermion 
at the time of the gauge field saturation, $k/a_\mathrm{sat}\simeq g' Q E_c \tau_\mathrm{prod}/a_\mathrm{sat}$, is now estimated 
as $\sim 0.14 g' Q H_I I(\xi_\mathrm{eff}) (a_\mathrm{sat}/a_\mathrm{end})$, 
which is larger than the Hubble scale as well as the coherence length of the hyperelectric fields
at the time of the saturation of the gauge field amplification for $\xi_\mathrm{eff}>1$, 
and hence the approximation of the homogeneous electric field is justified.
In the following discussion, we will take $\xi_{\rm eff} = 4$ achieved by $\xi_K\sim 10$ as a reference value for the investigation of the dynamics of the system. 

%%%%%%%%%%%%%%%%%%%
%%%%%%%%%%%%%%%%%%%%

%%%%%%%%%%%%%%%%%%%%%%%%%%%%%%%%%%%%%%%%
\subsection{Scattering and thermalization of produced charged fermions and reheating} \label{subsec_scat}
%%%%%%%%%%%%%%%%%%%%%%%%%%%%%%%%%%%%%%%%

In the previous subsection, we have seen how the pair-produced particles 
suppress the efficiency of the gauge field amplification. 
However, we have not investigated their thermalization, 
which would be important for reheating as well as the late-time evolution of the gauge fields.
In this subsection, we study the thermalization of the pair-produced particles and its effect on the dynamics of the background electric and magnetic fields
as well as the screening of the electric fields, with examining the time-evolution of the pair-produced particles. 

%%%%%%%%%%%%%%%%%%%%%%%%%%%%%%%%%%%%%%%%
\subsubsection{Non-thermalization of the pair-produced particles during the gauge field amplification}
%%%%%%%%%%%%%%%%%%%%%%%%%%%%%%%%%%%%%%%%

In this subsection, we shall see that the pair-produced particles are not thermalized until the saturation of the gauge field amplification. 
Let us first investigate the self-scattering of the LLL fermions and their (non-)thermalization. 
If the LLL fermions are accelerated by the constant and homogeneous electric field $E_\mathrm{eff}$ (together with $B_\mathrm{eff}$) for a sufficiently long period such that $s_i \gg B_\mathrm{eff}$, with $s_i$ denoting the center of mass energy squared of the fermion labeled by $i$, they are no longer confined along the magnetic field and their scattering rate is naively evaluated as
\begin{align}
    \Gamma_{\rm sc}^{\rm LLL} = \frac{g'^4}{12\pi s_i} n_{\psi_i}^{\rm LLL}. \label{standard_rate}
\end{align}
By taking the acceleration time as $\tau$ in the conformal time, their energy scale is evaluated as $s_i(\tau) = 2(g' q_i E_\mathrm{eff} a \tau)^2$, assuming that the acceleration is not disturbed by the scattering.
At the same time, the physical number density of the LLL fermions is evaluated by the anomaly equation (see Eq.~\eqref{nLLL}) as $n_{\psi_i}^\mathrm{LLL}(\tau) \sim (g'^2 q_i^2/4 \pi^2) E_\mathrm{eff} B_\mathrm{eff} a \tau$. 
The scatterings happen so often that the LLL fermions are thermalized if $\Gamma_\mathrm{sc}^\mathrm{LLL} a \tau \gg 1$. 
We find that with the above estimates, it is evaluated as
\begin{equation}
    \Gamma_{\rm sc}^{\rm LLL}   a \tau \simeq \frac{g'^4}{96 \pi^3} \ll 1,  \label{Gat_LLL}
\end{equation}
where we have used the condition $E_\mathrm{eff} = B_\mathrm{eff}$ (Eq.~\eqref{EMBR1}).
Apparently the LLL fermions are never thermalized. However, we need to take a special care in this situation. 
At the final stage of the gauge field amplification, the system is underoccupied or
the typical momentum of the LLL becomes much larger than the inverse of the mean separation length, 
which can also be expressed as the condition $T_{\rm wb}^{\mathrm{LLL}} < \omega_{\psi_i}^{\mathrm{LLL}}$, with $T_{\rm wb}^{\mathrm{LLL}} (\tau) \simeq \left((30/\pi^2 g_\ast) n^\mathrm{LLL}_{\psi_i} (\tau) \omega_{\psi_i}(\tau) \right)^{1/4}$
and $\omega_{\psi_i} = \sqrt{s_i/2}$ being the ``would-be temperature'' and the energy of the LLL fermion, respectively.
In such a stage the rate Eq.~\eqref{standard_rate} is not directly applied, 
where large-angle scatterings are assumed.
Instead, the main channel of the thermalization of the LLL fermions is turned out to be
the multiple small-angle soft gauge boson scatterings~\cite{Domcke:2018eki} (see also~\cite{Harigaya:2013vwa,Mukaida:2015ria,Ellis:2015jpg}), which needs to 
be evaluated by taking into account the Landau-Pomeranchuk-Migdal (LPM) process~\cite{Landau:1953um,Migdal:1956tc}. 
With these care, the thermalization rate (with non-Abelian gauge theories)
for $T_{\rm wb}^{\mathrm{LLL}} < \omega_{\psi_i}^{\mathrm{LLL}}$ case
is found to be given by~\cite{Gyulassy:1993hr,Arnold:2001ba,Arnold:2002ja,Kurkela:2011ti}
\begin{align}
    \Gamma_{\rm LPM}^{{\rm LLL}} (\tau) = 
        \frac{g'^4}{16\pi^2} T_{\rm wb}^{\mathrm{LLL}}  (\tau)  \sqrt{T_{\rm wb}^{\mathrm{LLL}} (\tau) /\omega_{\psi_i}^{\mathrm{LLL}} (\tau) }. \label{GLPM}
\end{align}
We find that  $\Gamma_{\rm LPM}^{{\rm LLL}} (\tau)  a \tau$ is an increasing function of $\tau$ and
\begin{equation}
    \left. \Gamma_{\rm LPM}^{{\rm LLL}} (\tau) a \tau \right|_{\tau = \eta_\mathrm{prod}} \simeq 1.3 \times 10^{-4}  \left(\frac{g_*}{106.75}\right)^{-\frac38} \left(\frac{g'}{0.3}\right)^{\frac{37}{8}} |q_i|^{\frac58} \left(\frac{ \mathcal{I}(\xi_{\rm eff})}{\mathcal{I}(4)}\right)^{\frac58} \left( \frac{a_{\rm sat}}{\frac{3\pi}{4}\sqrt{\xi_\mathrm{eff}} a_{\rm end}} \right)^{\frac52} \left(\frac{\xi_\mathrm{eff}}{4}\right)^{\frac54}, \label{GLPM_LLL}
\end{equation}
which is smaller than unity typically for $\xi_\mathrm{eff} < 13$. 
Note that such a large $\xi_\mathrm{eff}$ causes the too large backreaction as we have seen in Sec.~\ref{sec_gaugefieldproduction_backreaction} and is  
difficult to be obtained as can be seen in Fig.~\ref{fig:EMBR}. 
This suggests that the thermalization condition is not satisfied during the gauge field amplification for the reasonable value of $\xi_\mathrm{eff}$, even taking into account the LPM effect in the scattering rate. 
Therefore, we conclude that the LLL fermions would not be thermalized during the gauge field amplification. 

Next we examine the scattering of the HLL fermions.
In the situation of our interest, the inverse of the mean separation length of the HLL fermions is found to also be always much smaller than their typical momentum. 
Thus we shall evaluate their scattering rate with the LPM one, $\Gamma_\mathrm{LPM}^{\mathrm{HLL} (n)}$. 
In the same way {in} the case of the LLL fermions, we find that $\Gamma_\mathrm{LPM}^{\mathrm{HLL} (n)} a \tau$ is a {monotonically} increasing function {of time}, and for example, for $n=1$ HLL fermions, it is estimated as
\begin{equation}
    \left. \Gamma_{\rm LPM}^{{\rm HLL} (1)} (\tau) a \tau \right|_{\tau = \eta_\mathrm{prod}} \simeq 1.6 \times 10^{-5}  \left(\frac{g_*}{106.75}\right)^{-\frac38} \left(\frac{g'}{0.3}\right)^{\frac{37}{8}} |q_i|^{\frac58} \left(\frac{ \mathcal{I}(\xi_{\rm eff})}{\mathcal{I}(4)} \right)^{\frac58} \left( \frac{a_{\rm sat}}{\frac{3\pi}{4}\sqrt{\xi_\mathrm{eff}} a_{\rm end}} \right)^{\frac52} \left(\frac{\xi_\mathrm{eff}}{4}\right)^{\frac54}.
\end{equation}
Here we take the {typical} energy of the $n$-th HLL fermions {as} $\omega_{\psi_i}^{(n)} (\tau) = \sqrt{(g'|q_i|E_{\mathrm{eff}} a \tau)^2 + 2ng'|q_i|B_\mathrm{eff}}$ as indicated by the dispersion relation~\eqref{disp_rel}, and {the would-be temperature} {as} $T_{\rm wb}^{(n)} = \left((30/\pi^2 g_\ast) n^{(n)}_{\psi_i}(\tau)  \omega_{\psi_i}(\tau) \right)^{1/4}$ with the physical number density of the $n$-th HLL fermions $n^{(n)}_{\psi_i}(\tau) \simeq (g'^2 |q_i|^2/2 \pi^2) E_\mathrm{eff} B_\mathrm{eff} a \tau e^{-2 \pi n B_\mathrm{eff}/E_\mathrm{eff}}$.
We also find that $\Gamma_{\rm LPM}^{{\rm HLL} (n)} (\tau) a \tau$ at $\tau=\eta_\mathrm{prod}$ is even more suppressed for $n>1$ level.
Thus as long as $\xi_\mathrm{eff} < 14$, $\Gamma_{\rm LPM}^{{\rm HLL} (n)} (\tau) a \tau $ is always much smaller than the unity and it is unlikely that the HLL fermions are thermalized {before} the saturation of the gauge field amplification.
In other words, the validity of the estimate of the Schwinger effect in the previous subsection, where we do not take into account the scattering of the produced particles, is now confirmed.

%%%%%%%%%%%%%%%%%%%%%%%%%%%%%%%%%%%%
\subsubsection{Screening of the electric field}
%%%%%%%%%%%%%%%%%%%%%%%%%%%%%%%%%%%%

Next, let us investigate the screening of the (hyper) electric field
by the pair-produced particles.
After the gauge field amplification saturates, the energy injection from the background ALP into the hyper electric field  {becomes negligible}.
{As a consequence,} 
the hyper electric field begins to decay due to the screening by the pair-produced particles.
The screening by the charged particles is characterized by the electric conductivity. 
Strictly speaking, it would be desirable if we could evaluate it by using Kubo formula 
with taking into account the accurate phase space distribution, which is practically difficult. 
Before the thermalization (See Eq~\eqref{ath}), we instead evaluate it by adopting the Drude model.
The electric conductivity carried by the particles labeled by $i$ is evaluated as
\begin{align}
    \sigma_i \sim \frac{n_{\psi_i} g'^2}{\omega_{\psi_i} \Gamma_{\rm LPM}} = \frac{16\pi^2}{g'^2} \left( \frac{\pi^2 g_\ast}{30} \right)^{\frac38} n_{\psi_i}^{\frac58} \omega_{\psi_i}^{-\frac78}, \label{Drude}
\end{align}
where we take the characteristic time scale as the inverse of the LPM scattering rate (Eq.~\eqref{GLPM}).
Since the plasma is dominated by the LLL fermions, {by} substituting Eq.~\eqref{nLLLphys} and $\omega_{\psi_i}^{\rm LLL} = \sqrt{s_i/2}$ into Eq.~\eqref{Drude} {and taking into account the redshift}, 
{we} obtain
\begin{align}
    \sigma_i \sim 7.4 \times 10 H_I \left(\frac{g_*}{106.75}\right)^{\frac38} \left(\frac{g'}{0.3}\right)^{-\frac{13}{8}}  |q_i|^{\frac38} \left(\frac{{\cal I}(\xi_{\rm eff})}{{\cal I}(4)}\right)^{\frac{3}{8}} \left( \frac{a_{\rm sat}}{\frac{3\pi}{4}\sqrt{\xi_\mathrm{eff}} a_{\rm end}} \right)^{-\frac32} \left(\frac{\xi_\mathrm{eff}}{4}\right)^{\frac34} \left( \frac{a}{a_{\rm sat}} \right)^{-1}.
\end{align}
We can see that it is already much larger than the Hubble rate $H = H_I (a/a_{\rm end})^{-3}$ at $a = a_{\rm sat}$.
Since the electric conductivity $\sigma$ means that the electric fields are screened in a time scale of $\sigma^{-1}$, the hyper electric field is immediately screened {just} after the gauge field amplification saturates.
The energy of the hyper electric fields is converted to the one carried by the LLL fermions. 
This would change the estimate of their thermalization and reheating slightly, but quantitatively 
it is not significant in the parameter space we are interested in.
On the other hand, the hyper magnetic fields {are not screened}  
for a considerably long time $\sim \sigma \lambda_{\rm phys}^2$ with $\lambda_\mathrm{phys} \gg \sigma^{-1}$, 
and we assume that they evolve as long-range non-oscillatory stochastic fields after that.
Their evolution would affect the thermal history of the early Universe, {on which} we will {discuss} in the next section.

%%%%%%%%%%%%%%%%%%%%%%%%%%%%%%%%%%%%%%%%
\subsubsection{Eventual thermalization of the pair-produced particles and reheating}
%%%%%%%%%%%%%%%%%%%%%%%%%%%%%%%%%%%%%%%%

Let us now investigate how the produced particles are eventually thermalized
and dominate the energy density of the Universe.
The physical energy density of the LLL as well as the HLL fermions at the saturation of the gauge field amplification, $\tau = \eta_\mathrm{prod} = (2 a_\mathrm{sat}H_{\rm sat})^{-1}$ or $a= a_\mathrm{sat}$, is approximated as
\begin{align}
    \rho_\psi^{\rm LLL}(\eta_{\rm prod}) &= \sum_i \frac{1}{a_{\rm sat}^4 V} \int \df^3 x \int \frac{\df k_y \df k_z}{(2\pi)^2} [h_0(X)]^2 \Theta \left(-k_z\right) \Theta \left(k_z + g|q_i|E_c \eta_{\rm prod}\right) \nonumber \\
    &\qquad\qquad\qquad\qquad\qquad\qquad\qquad\qquad\qquad\quad \times \left(k_z + g'|q_i|E_c \eta_{\rm prod}\right) \nonumber \\ 
    &= \frac{g'^3}{4 \pi^2 a_{\rm sat}^4} Q_3 B_c E_c ^2 \eta_\mathrm{prod}^2
     \nonumber \\
    &\simeq 0.48 \times \left(\frac{{\cal I}(\xi_{\rm eff})}{{\cal I}(4)}\right)^3  \left(\frac{g'}{0.3} \right)^3 \left( \frac{Q_3}{41/12} \right) H_I^4, \label{rhoLLL} \\
 \rho_\psi^{(n)}(\eta_{\rm prod}) &= \sum_i \frac{2}{a_{\rm sat}^4 V} \int \df^3 x \int \frac{\df k_y \df k_z}{(2\pi)^2} [h_0(X)]^2 \Theta \left(-k_z\right) \Theta \left(k_z + g|q_i|E_c \eta_{\rm prod}\right) \notag \\
    &\qquad\qquad\qquad\qquad\qquad\qquad\qquad \quad \times \left(k_z + g'|q_i|E_c \eta_{\rm prod}\right) e^{-2 n \pi B_c/E_c} \nonumber \\ 
    &= \frac{g'^3}{2 \pi^2 a_{\rm sat}^4} Q_3 B_c E_c ^2 \eta_\mathrm{prod}^2 e^{-2 \pi n B_c/E_c}  \nonumber \\
    &\simeq 0.96 \times \left(\frac{{\cal I}(\xi_{\rm eff})}{{\cal I}(4)}\right)^3  \left(\frac{g'}{0.3} \right)^3 \left( \frac{Q_3}{41/12} \right) H_I^4e^{-2 \pi n }, \label{rhoHLL} 
\end{align}
where we have used $E_c = B_c = a^2 E_\mathrm{eff}=a^2 B_\mathrm{eff}$ and Eq.~\eqref{EMBR1}.
We can see that the energy density of the HLL fermions is suppressed by a factor of $e^{-2 \pi n}$ and the total energy density of the produced particles is dominated by the LLL fermions.
We also find that the energy density of the LLL fermions is typically larger than that of the electric and magnetic fields, 
\begin{align}
    \frac{\rho_\psi^{\rm LLL}}{\rho_{EE}+\rho_{BB}} = 
    2.6 \times \left(\frac{g'}{0.3} \right)^3  \left( \frac{Q_3}{41/12} \right) \left(\frac{{\cal I}(\xi_{\rm eff})}{{\cal I}(4)}\right) \left( \frac{a_{\rm sat}}{\frac{3\pi}{4}\sqrt{\xi_\mathrm{eff}} a_{\rm end}} \right)^4 \left(\frac{\xi_\mathrm{eff}}{4}\right)^2, \label{ratio_fer_EB}
\end{align}
where we have used Eqs.~\eqref{eqrhoEE}  and~\eqref{eqrhoBB} and taken $\xi_{\rm eff} = 4$ as a reference value. 

As has been discussed in the previous section, the hyperelectric fields are screened just after the 
saturation of gauge field amplification. 
Thus we assume that thereafter they evolve adiabatically, $n_\psi \propto a^{-3}$ and $\omega_\psi \propto a^{-1}$. 
Under this assumption, the LPM scattering rate evolves as $\Gamma_{\rm LPM}^{\rm LLL} \propto a^{-1}$  and eventually becomes larger than the Hubble rate, which is proportional to $a^{-3}$ during kination. 
Once it becomes $H \sim \Gamma_{\rm LPM}^{\rm LLL}$, we expect that the LLL fermions are thermalized.
The scale factor at that time is estimated as
\begin{align}
    a=a_{\rm th} = 3.3 \times 10^2 & \left(\frac{g_*}{106.75}\right)^{\frac{3}{16}} \left(\frac{g'}{0.3}\right)^{-\frac{37}{16}}  |q_i|^{-\frac{5}{16}} \notag \\
    &\times\left(\frac{{\cal I}(\xi_{\rm eff})}{{\cal I}(4)}\right)^{-\frac{5}{16}} \left( \frac{a_{\rm sat}}{\frac{3\pi}{4}\sqrt{\xi_\mathrm{eff}} a_{\rm end}} \right)^{-\frac14} \left(\frac{\xi_\mathrm{eff}}{4}\right)^{-{\frac18}} a_{\rm end}. \label{ath}
\end{align}
The relevant quantities at the time of LLL fermion thermalization are given as
\begin{align}
H_\mathrm{th} = \left(\frac{a_\mathrm{end}}{a_\mathrm{th}}\right)^3 H_I \simeq 2.8 \times 10^5 \mathrm{GeV}  &\left(\frac{g_*}{106.75}\right)^{-\frac{9}{16}} \left(\frac{g'}{0.3}\right)^{\frac{111}{16}} \left(\frac{{\cal I}(\xi_{\rm eff})}{{\cal I}(4)}\right)^{\frac{15}{16}}  \notag \\
& \times \left( \frac{a_{\rm sat}}{\frac{3\pi}{4}\sqrt{\xi_\mathrm{eff}} a_{\rm end}} \right)^{\frac34}  \left(\frac{\xi_\mathrm{eff}}{4}\right)^{\frac38}  \left(\frac{H_I}{10^{13} \mathrm{GeV}}\right), \label{hth} \\
T_\mathrm{th} = \left(\frac{30 \rho_\psi^\mathrm{LLL}(a_\mathrm{th})}{\pi^2 g_*}\right)^{1/4}
\simeq 4.9 \times 10^{10} \mathrm{GeV} &\left(\frac{g_*}{106.75}\right)^{-\frac{7}{16}} \left(\frac{g'}{0.3}\right)^{\frac{49}{16}}  \left(\frac{Q_3}{41/12}\right)^{1/4} \left(\frac{{\cal I}(\xi_{\rm eff})}{{\cal I}(4)}\right)^{\frac{17}{16}}  \notag \\
& \times \left( \frac{a_{\rm sat}}{\frac{3\pi}{4}\sqrt{\xi_\mathrm{eff}} a_{\rm end}} \right)^{\frac54}  \left(\frac{\xi_\mathrm{eff}}{4}\right)^{\frac58}  \left(\frac{H_I}{10^{13} \mathrm{GeV}}\right), \label{tth} \\ 
B_\mathrm{p,th} = B_\mathrm{eff} (a_\mathrm{th}) \simeq 8.6\times 10^{21} \mathrm{GeV}^2 &\left(\frac{g_*}{106.75}\right)^{-\frac{3}{8}} \left(\frac{g'}{0.3}\right)^{\frac{37}{8}}   \left(\frac{{\cal I}(\xi_{\rm eff})}{{\cal I}(4)}\right)^{\frac{13}{8}}  \notag \\
& \times \left( \frac{a_{\rm sat}}{\frac{3\pi}{4}\sqrt{\xi_\mathrm{eff}} a_{\rm end}} \right)^{\frac12}  \left(\frac{\xi_\mathrm{eff}}{4}\right)^{\frac14}  \left(\frac{H_I}{10^{13} \mathrm{GeV}}\right)^2, \label{bth} \\
\lambda_\mathrm{phys,th} = \lambda_\mathrm{phys}(a_\mathrm{th}) = 1.1 \times 10^{-10} \mathrm{GeV}^{-1}  &\left(\frac{g_*}{106.75}\right)^{\frac{3}{16}} \left(\frac{g'}{0.3}\right)^{-\frac{37}{16}}   \left(\frac{{\cal I}(\xi_{\rm eff})}{{\cal I}(4)}\right)^{-\frac{5}{16}}  \notag \\
& \times \left( \frac{a_{\rm sat}}{\frac{3\pi}{4}\sqrt{\xi_\mathrm{eff}} a_{\rm end}} \right)^{-\frac14}  \left(\frac{\xi_\mathrm{eff}}{4}\right)^{\frac78}  \left(\frac{H_I}{10^{13} \mathrm{GeV}}\right)^{-1}, \label{lth}
\end{align}
where we have taken $|q_i|=1$ to determine $a_\mathrm{th}$. 
Note that we here assume that the thermalization occurs during the kination era,
which is satisfied for the threshold $\xi_{\rm eff} \lesssim 6.6$ 
when $H_I=10^{13}$ GeV and for larger threshold of $\xi_\mathrm{eff}$ with
smaller $H_I$. For $\xi_\mathrm{eff}$ larger than the threshold,
radiation, mainly composed of the LLL fermions, 
dominates the Universe before 
thermalization. 
In this case, the scale factor at the thermalization is larger than the estimate of Eq.~\eqref{ath}.
As we can see from Fig.~\ref{fig:EMBR}, however, larger $\xi_\mathrm{eff}$ requires much larger $\xi_K$.
Such an inflation model is difficult to construct with avoiding the strong coupling problem or too large backreaction
problem. 
Hereafter we thus focus on the case for smaller $\xi_\mathrm{eff}$ than the threshold.

As mentioned, 
the energy density of the thermal LLL fermions eventually dominates the energy density of the Universe. 
The Hubble parameter and the (would-be) temperature at the time when the energy density of radiation, $\rho_\mathrm{rad}$,   
dominates the Universe (namely, reheating temperature) are given as
\begin{align}
H_\mathrm{re} &= 0.5 \mathrm{GeV}\left(\frac{g'}{0.3}\right)^{9/2} \left(\frac{Q_3}{41/12}\right)^{3/2}  \left(\frac{\mathcal{I}(\xi_\mathrm{eff})}{\mathcal{I}(4)}\right)^{9/2} \left(\frac{\xi_\mathrm{eff}}{4}\right)^3 \left(\frac{H_I}{10^{13} \mathrm{GeV}}\right)^4, \label{hre} \\
T_\mathrm{re} &=6 \times 10^8 \mathrm{GeV} \left(\frac{g_*}{106.75}\right)^{-1/4} \left(\frac{g'}{0.3}\right)^{9/4} \left(\frac{Q_3}{41/12}\right)^{3/4}  \left(\frac{\mathcal{I}(\xi_\mathrm{eff})}{\mathcal{I}(4)}\right)^{9/4} \left(\frac{\xi_\mathrm{eff}}{4}\right)^{3/2} \left(\frac{H_I}{10^{13} \mathrm{GeV}}\right)^2,
\end{align}
where we have used Eqs.~\eqref{EMBR1} and \eqref{rhoLLL}. 
Note that this estimate for the cosmic expansion 
does not depend on the condition if the LLL fermions are thermalized or not
(The ``would-be'' temperature is the temperature if the relativistic components are thermalized.).
Here we take $\rho_\mathrm{rad} = \rho_\psi^\mathrm{LLL}$ and omitted other relativistic components 
such as $\rho_\mathrm{EE}$ and $\rho_\mathrm{BB}$. 
As we can see from Eq.~\eqref{ratio_fer_EB}, they are comparable to the LLL fermion 
or can be a bit larger, but the estimate does not change much. 
Thus we conclude that  reheating well before the electroweak symmetry breaking or the BBN
is realized for sufficiently large Hubble parameter during inflation and 
appropriate value of the effective instability parameter $\xi_\mathrm{eff}$, 
which is much more efficient than the usual gravitational reheating~\cite{Parker:1969au,Zeldovich:1971mw,Ford:1986sy,Hashiba:2018iff}. 
For example, for $\xi_\mathrm{eff}=4$, reheating takes place before the electroweak symmetry breaking for 
$H_I>10^{10}$~GeV and before the BBN for  $H_I\gg10^{7}$~GeV. 
It can be also said that this is a realization of the Schwinger reheating~\cite{Tangarife:2017rgl} without a dark sector.

%%%%%%%%%%%%%%%%%%%%%%%%%%%%%%%%%%%%%%%%%%%%
\section{Late-time evolution of magnetic fields and cosmological consequences}
\label{sec_timeevolution}
%%%%%%%%%%%%%%%%%%%%%%%%%%%%%%%%%%%%%%%%%%%%

In the previous section we have seen that the fermions produced by the Schwinger effect
can be successfully thermalized and dominate the energy density of the Universe well 
before the electroweak symmetry breaking.
However, it is not enough to conclude that the Universe that experienced this mechanism is consistent with the 
present Universe, since the relics might cause some unwanted phenomena.
Namely, while we have seen that the
hyper electric fields are immediately damped, the hyper magnetic fields can last for a relatively long time, 
which should pass the constraints of the number of additional relativistic degrees of freedom especially from the 
Big Bang Nucleosynthesis (BBN)~\cite{1969Natur.223..938G,1970ApJ...160..451M,Cheng:1993kz,Grasso:1994ph,Kernan:1995bz,Cheng:1996yi,Kernan:1996ab} as well as the baryon overproduction/baryon isocurvature perturbation~\cite{Giovannini:1997gp,Giovannini:1997eg,Bamba:2006km,Fujita:2016igl,Kamada:2016eeb,Kamada:2016cnb,Kamada:2020bmb}. 
To check if the (hyper) magnetic fields are harmless or not, we should examine their evolution.
In this section, we examine the evolution of the magnetic fields with the magnetohydrodynamics (MHD), focusing on the former problem. 
Indeed, we will see that the baryon overproduction 
is unavoidable for the electroweak crossover in the light of the 125 GeV Higgs~\cite{Kamada:2016cnb}
unless the chiral plasma instability~\cite{Joyce:1997uy,Akamatsu:2013pjd,Schober:2017cdw,Kamada:2018tcs,Masada:2018swb} completely cancels the chirality and helicity~\cite{Domcke:2019mnd}. 
We discuss the way out of that in Sec.~\ref{sec_baryon}.

%%%%%%%%%%%%%%%%%%%%%%%%%%%%%%%%%%%%%%%%%%%%%%%%%%%%
\subsection{Evolution of the magnetic fields and the BBN constraint}
%%%%%%%%%%%%%%%%%%%%%%%%%%%%%%%%%%%%%%%%%%%%%%%%%%%%

After the saturation of the gauge field amplification, the (hyper) magnetic fields are diffused and damped exponentially by the conducting fluid in a relatively slow time scale $\sigma \lambda_\mathrm{phys}^2$, 
in the absence of the bulk velocity fields and the chiral magnetic effect (CME)~\cite{Vilenkin:1980fu,Fukushima:2008xe}.
Once the LLL fermions are thermalized, 
the electric conductivity gets to be evaluated by the one for the high-temperature thermal plasma,
$\sigma \simeq (1/g'^2 \ln g'^{-1}) T=c_\sigma T$ 
with $c_\sigma \simeq 10^2$ ~\cite{Baym:1997gq,Arnold:2000dr}. 
At the same time, the bulk velocity fields ${\bm v}$ are excited, obeying the Navier-Stokes equation,
\begin{equation}
\frac{\partial}{\partial \eta} {\bm v} + {\bm v} \cdot {\bm \nabla} {\bm v} = \nu_c {\bm \nabla}^2 {\bm v} +  \frac{1}{\rho_c+p_c}({\bm \nabla} \times {\bm B}_c) \times {\bm B}_c -  \frac{1}{\rho_c+p_c}{\bm \nabla} p,  \label{NSeq}
\end{equation} 
where $\rho_c$ and $p_c$ are the comoving energy density and pressure of the fluid, respectively, and $\nu_c \sim (1/g'^4 \ln g'^{-1})(aT)^3/(\rho_c+p_c) =c_\nu /(a T)$ with $c_\nu \simeq 10$~\cite{Arnold:2000dr,Durrer:2013pga} is the viscosity normalized by $\rho_c+p_c$. 
Note that we have come back to the comoving frame and assumed the incompressibility (${\bm \nabla} \cdot {\bm v} =0$). 
As a result, magnetic fields evolve according to the Maxwell equation with the MHD approximation, 
\begin{equation}
\frac{\partial}{\partial \eta} {\bm B}_c = \frac{1}{\sigma_c}{\bm \nabla}^2 {\bm B}_c + {\bm \nabla} \times ({\bm v}\times {\bm B}_c) +\frac{g'^2}{2\pi^2 \sigma_c} \mu_5 {\bm \nabla} \times {\bm B}_c,  \label{mxwl}
\end{equation}
where $\sigma_c \equiv a \sigma\simeq c_\sigma a T$ is the comoving electric conductivity
and the  last term is the CME induced current. 
If the velocity fields become strong enough, the magnetic fields no longer simply decay with the diffusion 
but enter the cascade regime. 
In this regime, magnetic fields do not decay exponentially but show a power-law decay. 
If the magnetic fields are maximally helical, they evolve according to the so-called inverse cascade.

Though the appropriate numerical MHD simulation is needed to determine precisely in which case the system
enters the cascade regime, it is useful to introduce the Reynolds numbers to give the criteria.
The magnetic Reynolds number is defined as
\begin{equation}
R_m \equiv \sigma_c \lambda_c v, 
\end{equation}
where $v$ is the typical amplitude of the velocity field. Here the quantities are defined in the comoving frame. 
It gives the typical ratio between advection term (second term of Eq.~\eqref{mxwl}) and the diffusion term 
(first term of Eq.~\eqref{mxwl}). 
We can set the criteria that the magnetic fields enter the cascade regime 
in an eddy-turnover time, $\sim \lambda_c/v$, 
if the magnetic Reynolds number is larger than unity so that the diffusion is less efficient\footnote{
    Here we assume that the CME (third term of Eq.~\eqref{mxwl}) is subdominant. 
    It can be important at a later time to cause the chiral plasma instability at smaller scales~\cite{Domcke:2019mnd}, 
    but it does not affect the estimate.
}.%%%%%%%%%%%%%%%%%%%%%%%%%% 

The typical amplitude of the velocity fields can be estimated by investigating the kinetic Reynolds number, defined as
\begin{equation}
R_e \equiv \frac{v \lambda_c}{\nu_c}.
\end{equation}
It gives the typical ratio between the advection term (second term of the left-hand side of Eq.~\eqref{NSeq}) 
and the diffusion term (first term of the right-hand side of Eq.~\eqref{NSeq}). 
If both the magnetic and kinetic Reynolds number is much larger than the unity, 
the advection term is comparable to the Lorentz force term (second term of the right-hand side of Eq.~\eqref{NSeq}), 
which suggests that the equipartition is reached, $\rho_c v^2/2 \simeq  B_c^2/2$. 
In this case, the evolution of the system is fully nonlinear and such a regime is called as the ``turbulence'' regime. 
Note that we always get $R_m \gg R_e$ from the relation between the electric conductivity and viscosity. 
On the other hand, for $R_m\gg 1 >R_e$, the advection term in Eq.~\eqref{NSeq} is negligible and 
the diffusion term would be balanced to the Lorenz force term, $\nu_c v/\lambda_c \simeq B_c^2/\rho_c$, or $\rho_c v^2/2 \simeq  R_e B_c^2/2$. 
This case is called ``viscous'' regime. 
To summarize, the rough estimate of the velocity fields excited by the magnetic fields is given as
\begin{equation}
v \simeq \left\{ \begin{array}{ll}
\dfrac{B_c}{\sqrt{\rho_c}} = \dfrac{B_\mathrm{p}}{\sqrt{\rho_\mathrm{p}}} =\left(\dfrac{30}{\pi^2 g_*}\right)^{1/2} \dfrac{B_\mathrm{p}}{T^2}, & \text{for} \quad R_m \gg R_e >1,\\ \\
\dfrac{B_c^2 \lambda_c}{\nu_c (\rho_c+p_c)} =\dfrac{B_\mathrm{p}^2 \lambda_\mathrm{phys}}{\nu_\mathrm{p} (\rho_\mathrm{p}+p_\mathrm{p})} \sim \dfrac{3}{\pi^2 g_*} \left(\dfrac{c_\nu}{10}\right)^{-1} \dfrac{B_\mathrm{p}^2 \lambda_\mathrm{phys}}{T^3},& \text{for} \quad R_m \gg 1>R_e,
\end{array}\right. \label{velre}
\end{equation}
where we have rewritten in terms of the physical quantities.
From the helicity conservation, $h_c \sim B_c^2 \lambda_c = \mathrm{const}.$ 
and the condition that the coherence length of the magnetic fields is roughly equal to the eddy turnover scale 
$\sim v \eta$ (or by solving the Maxwell equation (Eq.~\eqref{mxwl}) only with the  advection term~\cite{Domcke:2019mnd}), 
we obtain the cascade law as
\begin{align}
B_c \propto \eta^{-1/3}, \quad \lambda_c \propto \eta^{2/3}, \quad v \propto \eta^{-1/3}, \quad \text{for} \quad R_m \gg R_e >1,\\
B_c \propto \eta^{-1/2}, \quad \lambda_c \propto \eta, \quad v = \mathrm{const.}, \quad \text{for} \quad R_m \gg 1>R_e, \label{icvis}
\end{align}
after the eddy-turnover time. 
Note that here we have assumed that the chiral plasma instability does not occur. 
If it takes place to erase total magnetic helicity, the decay of the magnetic fields can be stronger. 
In our present purpose it is enough not to consider such a case, 
because the case without this phenomena is more problematic for the BBN due to larger amount of remaining 
magnetic fields. 

Now we can examine the evolution of the system of our interest by evaluating
the magnetic and kinetic Reynolds number. 
Evaluating the parameters at the time of the LLL thermalization, 
from Eqs.~\eqref{tth}, \eqref{bth}, \eqref{lth}, and \eqref{velre}, we find that the kinetic Reynolds number is smaller than the unity for $\xi_\mathrm{eff} \lesssim 7$ 
as 
\begin{align}
R_e =& \frac{1}{\rho_c+p_c} \left(\frac{B_c \lambda_c}{\nu_c}\right)^2 = \frac{30}{\pi^2 g_*}  \left(\frac{B_\mathrm{p} \lambda_\mathrm{p}}{c_\nu T}\right)^2  \notag \\
\simeq & 1 \times 10^{-1} \left(\frac{g_\ast}{106.75} \right)^{-\frac12} \left(\frac{g'}{0.3} \right)^{-\frac32} \left(\frac{c_\nu}{10}\right)^{-2} \notag \\
&\times \left( \frac{Q_3}{41/12} \right)^{-\frac14} \left(\frac{{\cal I}(\xi_{\rm eff})}{{\cal I}(4)}\right)^{\frac12} \left( \frac{a_{\rm sat}}{\frac{3\pi}{4}\sqrt{\xi_\mathrm{eff}} a_{\rm end}} \right)^{-2} \left(\frac{\xi_\mathrm{eff}}{4}\right). 
\end{align}
Then the magnetic Reynolds number is evaluated as
\begin{align}
R_m = & \sigma_c \frac{B_c^2 \lambda_c^2}{\nu_c (\rho_c+p_c)}  = \frac{30}{\pi^2 g_*} \frac{c_\sigma}{c_\nu}  \left(\frac{B_\mathrm{p} \lambda_\mathrm{p}}{c_\nu T}\right)^2 \notag \\
\simeq &1 \times 10^2 \left(\frac{g_\ast}{106.75} \right)^{-\frac12} \left(\frac{g'}{0.3} \right)^{-\frac32} \left(\frac{c_\nu}{10}\right)^{-3} \left(\frac{c_\sigma}{10^2}\right) \notag \\
&\times \left( \frac{Q_3}{41/12} \right)^{-\frac14} \left(\frac{{\cal I}(\xi_{\rm eff})}{{\cal I}(4)}\right)^{\frac12} \left( \frac{a_{\rm sat}}{\frac{3\pi}{4}\sqrt{\xi_\mathrm{eff}} a_{\rm end}} \right)^{-2} \left(\frac{\xi_\mathrm{eff}}{4}\right), 
\end{align}
which is larger than the unity for $\xi_\mathrm{eff}>1$. 
Thus we determine that in the parameters of our interest, $1 \lesssim \xi_\mathrm{eff} \lesssim 6.6$, 
the magnetic fields would evolve with the inverse cascade in the viscous regime (Eq.~\eqref{icvis}). 
The eddy-turnover time (in the comoving frame) is evaluated as
\begin{align}
\sigma_c \lambda_c^2 \simeq  5.7 \times 10^{-8} \mathrm{GeV}^{-1} &a^{-1} \left(\frac{g_\ast}{106.75} \right)^{-\frac{1}{16}} \left(\frac{g'}{0.3} \right)^{-\frac{25}{16}}  \left(\frac{c_\sigma}{10^2}\right) \left( \frac{Q_3}{41/12} \right)^{\frac14} \notag \\
&\times \left(\frac{{\cal I}(\xi_{\rm eff})}{{\cal I}(4)}\right)^{\frac{9}{16}} \left( \frac{a_{\rm sat}}{\frac{3\pi}{4}\sqrt{\xi_\mathrm{eff}} a_{\rm end}} \right)^{\frac34} \left(\frac{\xi_\mathrm{eff}}{4}\right)^{\frac{11}{8}} \left(\frac{H_I}{10^{13}\mathrm{GeV}}\right)^{-1}, 
\end{align}
which suggests that the magnetic fields starts to evolve according to the inverse cascade at
\begin{align}
H=H_\mathrm{IC} \simeq 0.9\times 10^7\mathrm{GeV} &\left(\frac{g_\ast}{106.75} \right)^{\frac{1}{16}} \left(\frac{g'}{0.3} \right)^{\frac{25}{16}}  \left(\frac{c_\sigma}{10^2}\right)^{-1} \left( \frac{Q_3}{41/12} \right)^{-\frac14} \notag \\
&\times \left(\frac{{\cal I}(\xi_{\rm eff})}{{\cal I}(4)}\right)^{-\frac{9}{16}} \left( \frac{a_{\rm sat}}{\frac{3\pi}{4}\sqrt{\xi_\mathrm{eff}} a_{\rm end}} \right)^{-\frac34} \left(\frac{\xi_\mathrm{eff}}{4}\right)^{-\frac{11}{8}} \left(\frac{H_I}{10^{13}\mathrm{GeV}}\right). \label{hic}
\end{align}

With the estimation in the above, we can examine the constraints on the abundance of the magnetic fields
from the BBN, which gives the upper bound of the magnetic fields. 
Since the magnetic fields act as additional relativistic degrees of freedom, 
their abundance is characterized by the effective number of neutrino flavors, $N_\mathrm{eff}$, 
as
\begin{equation}
\frac{\rho_{BB}}{\rho_\mathrm{rad}} = \frac{7 \Delta N_\mathrm{eff}}{22+7N_\mathrm{eff}}
\end{equation}
with $\Delta N_\mathrm{eff} \equiv N_\mathrm{eff}- 3$. Recent constraints on $N_\mathrm{eff}$ 
from the BBN~\cite{Fields:2019pfx} tells $\Delta N_\mathrm{eff} < 0.16$. 
Thus the energy density of magnetic fields at the BBN should be constrained as
\begin{equation}
\frac{\rho_{BB}}{\rho_\mathrm{rad}}  < 2.5 \times 10^{-2}. 
\end{equation}
Although at the time of the LLL thermalization, the radiation and magnetic energy density is the same order, 
the latter becomes to decay faster than the former once they enter the cascade regime. 
The ratio between the energy density of magnetic fields and that of the radiation in the viscous regime decays in proportion to $\eta^{-1}$ and hence $H^{2/3}$  during the kination era and $\eta^{-1}$ and hence $H^{1/2}$ during the radiation dominated era.
Comparing to the Hubble parameters at the onset of the cascade (Eq.~\eqref{hic}), reheating (Eq.~\eqref{hre}), 
and the BBN ($H\sim 10^{-24}$ GeV), 
we find that the duration of the cascade evolution of the magnetic fields are long enough
to satisfy the constraints from the BBN on the magnetic field energy density.

%%%%%%%%%%%%%%%%%%%%%%%%%%%%%%%%%%%%%%%%%%%%%%%%%%%%%%%%%%%%
\subsection{Comment on baryogenesis} \label{sec_baryon}
%%%%%%%%%%%%%%%%%%%%%%%%%%%%%%%%%%%%%%%%%%%%%%%%%%%%%%%%%%%%

Since the magnetic fields are generated before the electroweak symmetry breaking as the hypermagnetic fields
and maximally helical, baryon asymmetry is generated at the time of the electroweak symmetry breaking~\cite{Giovannini:1997gp,Giovannini:1997eg,Bamba:2006km,Fujita:2016igl,Kamada:2016eeb,Kamada:2016cnb}. 
As has been shown in Ref.~\cite{Kamada:2016cnb} (see also Ref.~\cite{Kamada:2020bmb} for the simple derivation), 
the resultant baryon-to-entropy ratio is evaluated as
\begin{equation}
\eta_B \equiv \frac{n_B}{s} \sim 4 \times 10^{-4} \frac{h_c}{s_c}, 
\end{equation}
where $s_c = (4 \pi^2/45) g_{*s} (aT)^3$ is the comoving entropy density. 
Since the helicity ($h_c = \lambda_c B_c^2$) is conserved during the cascade process, 
one can see that 
the resultant baryon asymmetry $\eta_B$ is inevitably much larger than the observed value, 
$\eta_B \simeq 9 \times 10^{-11}$. 
Note that from Eqs.~\eqref{tth}, \eqref{bth}, and \eqref{lth}, we find $h_c/s_c \simeq 0.7 ({\cal I}(\xi_\mathrm{eff})/{\cal I}(4))^{-1/4} (\xi_\mathrm{eff}/4)^{3/8}$, 
unless the chiral plasma instability completely erases the hypermagnetic helicity and chirality 
well before the electroweak symmetry breaking. 
One might think that if the chiral plasma instability erases the helicity, the magnetic fields are harmless. 
However, even if it is the case so that the net baryon asymmetry vanishes, 
the magnetic fields would remain and they are inconsistent with the constraint from the inohomogeneous BBN
from the baryon isocurvature perturbation generated from the hypermagnetic fields~\cite{Kamada:2020bmb}. 
Therefore, in the present setup with the standard nature of the electroweak symmetry breaking, 
the baryon overproduction or too large baryon isocurvature perturbation is unavoidable, 
and the resultant Universe is inconsistent with ours.

A way out from this constraint is to introduce additional entropy production 
by, {\it e.g.}, gravitational heavy particle production~\cite{Hashiba:2018iff}. 
However, in a simple model construction, it is difficult to have a sufficiently large entropy production 
due to the ineffectiveness of the gravitational particle production. 
In this case, the gravitational particle production is responsible for the reheating and it is no longer 
the realization of the Schwinger reheating. 
Another way out from this constraint is to assume the change of the electroweak symmetry breaking. 
The reason why we have an efficient conversion from the magnetic helicity to the baryon asymmetry
is that the sphaleron freezeout takes place earlier than the completion of the electroweak symmetry breaking
in terms of the effective weak mixing angle~\cite{DOnofrio:2014rug,DOnofrio:2015gop}. 
If a new physics lies just above the electroweak scale to modify the nature of the electroweak symmetry breaking
so that the change of the effective weak mixing angle 
takes place much earlier than the sphaleron freezeout, 
the resultant baryon asymmetry can be suppressed, in a similar way studied in Ref.~\cite{Kamada:2016eeb}. 
In such a case, if the cascade decay lasts until recombination,
the magnetic fields would remain as the intergalactic magnetic fields today. 
Since they evolve with the MHD cascade, their coherence length would be relatively small, 
but their field strength is relatively large so that they might be able to explain 
the blazar observations~\cite{Neronov:2010gir,Tavecchio:2010mk,Dolag:2010ni,Finke:2015ona,Fermi-LAT:2018jdy}. 
The late-time magnetic field evolution, especially with the chiral magnetic effect, 
is extremely complicated, and the precise estimate is left for future study.

%%%%%%%%%%%%%%%%%%%%%%%%%%%%%%%%%%%%%%%%%%%%
\section{Conclusion and Discussion}    
\label{sec_conclusion}
%%%%%%%%%%%%%%%%%%%%%%%%%%%%%%%%%%%%%%%%%%%%

Graceful exit in inflationary models where the inflaton continuously runs away  
after the inflationary phase, of which period is known as kination, is one of the most important issue for their realistic model building, 
albeit their phenomenologically interesting features. 
In such models, a coherent inflaton oscillation phase does not follow the inflationary era, 
and reheating of the Universe cannot be achieved by the inflaton particle decay.
In this article, we pointed out that the Chern-Simons coupling between the inflaton and the gauge fields is allowed
by the shift symmetry behind the model with kination, if we identify the inflaton is an ALP. 
We examined the U(1) gauge field production through 
the tachyonic instability during kination caused by the Chern-Simons coupling
to see if it can lead to successful reheating in these models.

First we investigated the dynamics of the U(1) gauge fields with the Chern-Simons coupling 
during ALP kination without other fields.
We found that the energy density of the gauge fields are enhanced by a factor of $\exp[\pi \xi_K]/\xi_K^3$ 
with $\xi_K (>1)$ being the instability parameter, similar to the gauge field amplification during inflation~\cite{Turner:1987bw,Garretson:1992vt,Anber:2006xt}, 
which is much more efficient than the gravitational particle production~\cite{Parker:1969au,Zeldovich:1971mw}. 
Since the sign of the ALP velocity does not change during the process,
only one helicity mode is amplified and the generated gauge fields are completely helical.  
While the gauge fields are amplified when the mode exited the horizon in the case of the one during inflation, 
they are once more amplified when they reenter the horizon (the mode should have exited the horizon once during inflation). 
Unless $\xi_K$ is very large (as large as 10), the energy density of the gauge fields are much smaller than that of 
the kinetic energy density of the inflaton at the time of the saturation of their amplification even without considering the backreaction, 
but they eventually dominate it since the kinetic energy of the inflaton decreases much faster than that of gauge fields, $\rho_\mathrm{kin} \propto a^{-6}$.  
The results obtained here are not limited to the U(1) gauge field in the SM but also can be applied 
for any dark U(1) gauge fields. 

We then took into account the charged fermions with the SM particle species in mind and 
examined the Schwinger pair-particle production and its backreaction on the gauge field dynamics. 
In principle, the consistent treatment of the Schwinger effect in the dynamical gauge fields 
and its backreaction on the gauge field dynamics is a conceptually difficult problem, 
and a precise estimate is almost impossible at the best of our knowledge and technique. 
In this regard, we here adopted the way to give its rough estimate proposed in Ref.~\cite{Domcke:2018eki}, 
where the backreaction is characterized by the effective instability parameter $\xi_\mathrm{eff}$
obtained from the consistency equation for the original instability parameter $\xi_K$. 
In this treatment, the amplified gauge field spectrum as well as that of the pair-produced particles are 
described by $\xi_\mathrm{eff}$. 
The backreaction from the Schwinger effect becomes significant for $\xi_K \gtrsim 2$. 
We found that the energy density of the pair-produced (LLL) fermions are typically comparable to or larger than 
that of the gauge fields and the LLL fermions is thermalized well before they dominate the energy density of the Universe
for $\xi_\mathrm{eff} \lesssim 6.6$. 
The electric fields are screened soon after the saturation of the gauge field amplification. 
For sufficiently large Hubble parameter during inflation, the reheating of the Universe (when the energy density of 
radiation dominates over that of the inflaton) occurs well before the BBN and also the electroweak symmetry breaking
for $2 \lesssim \xi_\mathrm{eff} \lesssim 6$. 
Note that $\xi_\mathrm{eff}\simeq 4-5$ corresponds to $\xi_K\simeq {\cal O}(10)$, as can be seen in Fig.~\ref{fig:EMBR}.   
This process is a concrete realization of the ``Schwinger reheating''. 
If the magnetic fields survive for a too long time, it acts as additional relativistic degrees of freedom at the BBN, 
which spoils its success,
but we confirmed that the MHD cascade decay is sufficiently efficient to decrease the energy density of the 
magnetic fields before the BBN.

Our results strongly suggests that the reheating of the Universe 
can be successfully achieved in the inflation models followed by the kination era 
with the aid of the tachyonic instability of the gauge field caused by the Chern-Simons term
and the Schwinger effect. 
However, our analytic treatment, which is based on the one developed in Ref.~\cite{Domcke:2018eki}, 
relies on several approximations and assumptions. Namely, 
\begin{enumerate}
\item We adopted the Schwinger effect for the static and homogeneous gauge fields, while they have finite coherence
length and time dependence in our setup. 
\item We estimated the backreaction on the amplification of the gauge fields 
by requiring the consistency condition for their evolution equation but did not solve it simultaneously with the
evolution equation for fermions.  
\end{enumerate}
We admit that the estimate is much less precise than that for the inflationary magnetogenesis~\cite{Turner:1987bw,Garretson:1992vt,Anber:2006xt,Jimenez:2017cdr} 
where the system is expected to reach a static configuration eventually. 
Nevertheless, at the present knowledge and technique, our treatment would be the optimal one. 
To give more precise estimate, however, the technique to solve the dynamical co-evolution of the 
gauge fields and charged particles should be developed further.
We also assumed a simplified background cosmic expansion, with a instant connection between the 
inflation (with a constant Hubble parameter) and kination to make the analytic calculations possible. 
However, as has also been discussed in Ref.~\cite{Hashiba:2018iff} for the gravitational particle production, 
a smooth connection between the inflationary phase and the kination phase can make a 
difference in the gauge field amplification (especially for high momentum modes) and the time of reheating to a certain extent. 
Numerical studies with a concrete model of the background dynamics of inflation and the following kination era
are essential to compare the mechanism with the cosmological observations, 
which strongly depends on the time when the Universe becomes radiation dominated, 
and to identify the inflation model. 
Nevertheless, we believe that the investigation in the present study is the first step
for the concrete realization of the application of the Schwinger effect associated with 
the inflationary magnetogenesis to the reheating of the Universe, worth investigating in more depth.

%%%%%%%%%%%%%%%%%%%%%%%%%%%%%%%%%%%%%%%%%%%%
\section*{Acknowledgements}
%%%%%%%%%%%%%%%%%%%%%%%%%%%%%%%%%%%%%%%%%%%%
\small\noindent
The authors are grateful to Tomohiro Fujita and Kyohei Mukaida for useful discussions. 
This work is supported by JSPS KAKENHI, Grant-in-Aid for JSPS Fellows Nos.~20J10176 (S.\,H.), 19J21974 (H.\,N.), 
for Scientific Research (C) JP19K03842 (K.\,K.), 
and for Scientific Research on Innovative Areas 19H04610 (K.\,K.). 
S.\,H.~and H.\,N.~are supported by the Advanced Leading Graduate Course for Photon Science (ALPS).

%%%%%%%%%%%%%%%%%%%%%%%%%%%%%%%%%%%%%%%%%%%%%%%%%%%%%%%%%%%
%%%%%%%%    References 
\small
\bibliographystyle{utphys}
\bibliography{Ref.bib}%bib name can contain "_" , but not "/" 
\end{document}